\documentclass[twocolumn]{autart}
\usepackage{amsmath,amsfonts,amssymb}

\usepackage{algorithmic}
\usepackage{algorithm}
\usepackage{graphicx,cite}
\usepackage{mathrsfs,bm}
\usepackage{algorithm,algorithmic}

\usepackage{url}
\usepackage{booktabs,multirow}
\usepackage{mathtools}
\usepackage[subrefformat=parens]{subcaption}

\newcommand{\Real}[1]{{\rm Re}\,(#1)}
\newcommand{\Imag}[1]{{\rm Im}\,(#1)}
\newcommand{\trace}{{\rm tr}}

\newcommand{\mc}[1]{\mathcal{#1}}

\mathtoolsset{showonlyrefs}
\newcommand{\innerproduct}[2]{\langle #1, #2 \rangle_{\rm F}}

\newcommand{\stp}{\strut \partial}
\newcommand{\vecf}[1]{{\rm vec}\left(#1\right)}
\newcommand{\vecT}[1]{({\rm vec}(#1))^{\sf T}}

\newcommand{\iGam}{{\it \Gamma}}
\newcommand{\PDel}{P_{\Delta}}
\newcommand{\KDel}{K_{\Delta}}
\newcommand{\GDel}{G_{\Delta}}
\newcommand{\MDel}{M_{\Delta}}

\usepackage{marginnote,color,xcolor}

\begin{document}

\setlength{\abovedisplayskip}{5pt}
\setlength{\belowdisplayskip}{5pt}
\setlength{\abovedisplayshortskip}{2pt}
\setlength{\belowdisplayshortskip}{2pt}
\renewcommand{\arraystretch}{0.8}
\clubpenalty=50
\widowpenalty=50
\displaywidowpenalty=50
\brokenpenalty=50
\setlength{\abovecaptionskip}{2pt}
\setlength{\belowcaptionskip}{0pt}
\linespread{0.95}
\setlength{\parskip}{5pt}
\setlength{\parindent}{0em}

\begin{frontmatter}

\title{Adversarial Destabilization Attacks\\ to Direct Data-Driven Control\thanksref{footnoteinfo}}

\thanks[footnoteinfo]{This paper was not presented at any IFAC 
meeting. Corresponding author H. Sasahara.}

\author[HS]{Hampei Sasahara}\ead{hsasahara@g.ecc.u-tokyo.ac.jp}

\address[HS]{Graduate School of Information Science and Technology, the University of Tokyo, Tokyo 113-8656, Japan}


\begin{abstract}
This study explores the vulnerability of direct data driven control, particularly in the linear quadratic regulator (LQR) problem, to adversarial perturbations in offline collected data.
We focus on stealthy attacks that subtly alter training data to destabilize the closed-loop system while evading detection.
To craft such attacks, we propose Directed Gradient Sign Method (DGSM) and its iterative variant (I-DGSM), which adapt techniques from adversarial machine learning to align perturbations with the gradient of the closed-loop spectral radius.
A key technical contribution is an efficient and exact gradient computation method using implicit differentiation through the Karush-Kuhn-Tucker conditions of the underlying semidefinite program.
For defense, we introduce two strategies: (i) regularization to reduce controller sensitivity, and (ii) robust data-driven control that ensures stability under bounded perturbations.
Experiments across benchmark systems reveal that even imperceptibly small perturbations, up to ten times smaller than random noise, can lead to instability, while the proposed defenses significantly reduce attack success rates with minimal performance loss.
We also assess transferability under partial knowledge, demonstrating the importance of protecting training data.
This work highlights critical security risks in data driven control and proposes practical methods for both attack and defense.
\end{abstract}

\end{frontmatter}

\section{Introduction}
Cyber-physical systems (CPS), such as autonomous vehicles, consist of multiple interconnected layers, among which the perception and decision-making layer and the control layer are fundamental~\cite{Schwarting2018Planning}.
The perception and decision-making layer handles high-level interpretation, planning, and strategic decisions based on processed data, often leveraging artificial intelligence and machine learning techniques.
In contrast, the control layer is responsible for low-level closed-loop feedback control that directly governs physical system dynamics to ensure stability and performance.
It is well established that the perception and decision-making layer is vulnerable to various adversarial attacks that manipulate input data, leading to incorrect interpretations and potentially unsafe outcomes~\cite{Bruna2014Intriguing,Goodfellow2015Explaining,Carlini2016Hidden,Eykholt2018Robust,Tian2021Joint,Wang2021Stop,Deka2022Dynamically,Song2023Discovering}.
By contrast, the control layer has traditionally been viewed as more secure due to its reliance on well-defined physical models.

However, recent advances in data-driven control paradigms have significantly shifted this perspective~\cite{O2022Data,Dorfler2023Data,Schmitt2023Data}.
Data-driven control methods design controllers directly from observed input-output data without explicit system identification, enabling flexible and adaptive control even when precise models are unavailable~\cite{Campi2002Virtual,Kaneko2013Data,Lewis2013Reinforcement}.
This growing reliance on data at the control layer introduces new attack surfaces: if an adversary can manipulate the data used for controller synthesis, they may induce instability or degrade performance in ways previously unconsidered.
Consequently, the vulnerability of the control layer to data manipulation has become a critical concern.
Despite its importance, this area remains underexplored compared to the perception layer's adversarial risks with only a few recent studies addressing this issue~\cite{Yu2023Poisoning,Ikezaki2024Poisoning,Anand2025Analysis}.
Addressing vulnerabilities especially focusing on the emerging threats posed by data-driven control is essential for ensuring the comprehensive security and resilience of modern CPS.

In this study, we investigate the vulnerability of direct data-driven control~\cite{Dorfler2023Data} to adversarial perturbations.
Specifically, we consider an attacker who seeks to destabilize a closed-loop system by making small, intentional modifications to the input-output data used for controller synthesis.
We focus on the linear quadratic regulator (LQR) problem, a widely used standard benchmark control algorithm, formulated as a semidefinite program (SDP)~\cite{De2019Formulas}.
To assess worst-case risks, we assume a powerful white-box adversary with full knowledge of the system dynamics, the controller design algorithm, and the clean data.
As a specific adversarial strategy, we propose the directed gradient sign method (DGSM), an adaptation of the fast gradient sign method (FGSM) originally developed for attacking neural networks~\cite{Goodfellow2015Explaining}.
DGSM computes perturbations aligned with the gradient of the spectral radius (or eigenvalues) of the closed-loop system matrix, aiming to reduce system stability.
We further extend this approach with the iterative DGSM (I-DGSM), formulated as a projected gradient descent method, which refines the perturbations through multiple steps to increase their effectiveness.

A key challenge in applying I-DGSM lies in computing the gradient of the spectral radius with respect to the data.
Standard numerical differentiation, such as the central difference method, would require solving an SDP multiple times proportional to the data dimension, leading to significant computational costs.
To overcome this, we develop a novel approach based on implicit differentiation through the Karush-Kuhn-Tucker (KKT) conditions of the SDP, which requires solving the semidefinite program only once.
This leads to exact gradient computation at a significantly reduced cost, enabling efficient generation of adversarial perturbations.

Furthermore, in response to these vulnerabilities, we propose two defense methods: a regularization-based approach and a robust data-driven control approach.
Although the regularization has been originally introduced to mitigate random disturbances~\cite{Dorfler2022On,Dorfler2023Bridging}, we show both theoretically and numerically that it is also effective against adversarial attacks.
The robust data-driven control approach builds on the recent work~\cite{Kaminaga2025Data,Kaminaga2025DataACC} and guarantees closed-loop stability under bounded perturbations defined by a matrix ellipsoid.
This robustness framework provides theoretical assurances critical for deploying data-driven controllers in safety-critical settings.

To validate the severity of adversarial risks, we conduct extensive simulations on benchmark linear time-invariant systems, including five standard mechanical and aerospace models from the tutorials on control engineering~\cite{William1999Control,Dawn2025Tilbury}, as well as a triple-tank system representing a typical process control application~\cite{Blanke2016Diagnosis,Milosevic2019Estimating}.
A representative example with the triple tank system shows that perturbations with magnitudes as small as 0.1\% to 0.5\% of the data amplitude can induce instability while remaining visually indistinguishable from clean data.
Across all tested systems, our results consistently demonstrate that adversarial perturbations with magnitudes approximately one-tenth or less than those of uniformly distributed random noise are sufficient to destabilize the closed-loop system.
Furthermore, the proposed implicit differentiation method improves computational efficiency dramatically: compared to numerical differentiation, it achieves speedups of around 70 times, with further gains, up to nearly 3 times, when analytical derivatives are used.
In addition,
the proposed defense strategies significantly reduce the attack success rate.
Specifically, the regularization-based approach decreases the attack success rate from near 100 \% to as low as 5\% or below in several benchmark systems while maintaining control performance close to the nominal optimal.
Meanwhile, the robust data-driven control method guarantees closed-loop stability under bounded perturbations and achieves nominal performance for small perturbation levels, although its control performance degrades as the allowable perturbation size increases.
Finally, beyond worst-case white-box attacks, we explore effectiveness in more realistic adversarial scenarios involving limited attacker knowledge, referred to as data transferability.
Numerical results indicate that data transferability is moderate under the threat model and suggest that maintaining the confidentiality of the training data can serve as a practical defense against adversarial perturbations in control systems.

This work makes the following key contributions.
First, we reveal a fundamental vulnerability of direct data-driven LQR control methods to adversarial perturbations in the training data.
Second, we develop efficient perturbation algorithms, DGSM and its iterative variant I-DGSM, tailored to destabilize the resulting closed-loop system by exploiting the structure of the underlying SDP.
Third, we introduce an efficient attack synthesis approach using implicit differentiation through the KKT conditions of the control design problem.
Fourth, we propose two proactive defense strategies: a regularization-based method and a robust data-driven control approach.
Finally, we validate our methods through extensive numerical experiments demonstrating both the severity of the attacks and the effectiveness of the proposed defenses, including transferability analysis under partial knowledge scenarios.

Existing studies on attacks against data-driven control address different objectives and controller classes.
Specifically,~\cite{Yu2023Poisoning} investigates attacks on data-driven predictive control under abstract adversarial objectives.
The work~\cite{Ikezaki2024Poisoning} considers attacks on parametric controllers such as proportional-integral-derivative (PID) controllers designed via virtual internal model tuning (VIMT).
Further, in~\cite{Anand2025Analysis}, they focus on inducing a desired controller or degrading performance via bias injection in data-driven LQR.
While prior work on attacks against data-driven LQR primarily focuses on performance degradation, destabilizing attacks have not been systematically studied in this setting.
Since stabilization is a fundamental requirement in control systems, understanding adversarial impacts on this property is of primary importance.
Moreover, LQR serves as a canonical benchmark in control theory, making the study of its vulnerability to destabilizing attacks both practically and theoretically significant.
Beyond the attack analysis, we further propose two systematic defense mechanisms based on regularization and robust optimization, providing principled countermeasures that have not been addressed in prior work.

Preliminary versions of this work have been presented in~\cite{Sasahara2023Adversarial} and~\cite{Kaminaga2024Adversarial}.
In~\cite{Sasahara2023Adversarial}, the DGSM algorithm, a fundamental method for generating adversarial perturbations, was introduced.
An extension of this method, referred to as I-DGSM, was proposed in~\cite{Kaminaga2024Adversarial}.
While prior works rely on numerical differentiation, the present work proposes a more efficient gradient computation method, as detailed in Sec.~\ref{sec:alg}.
Furthermore, we introduce a new defense method based on robust data-driven control in Sec.~\ref{sec:def}.
Lastly, our numerical evaluation in Sec.~\ref{sec:exp} includes a broader and more diverse set of benchmark systems, providing a more comprehensive validation of the proposed methods.

\subsection*{Organization and Notation}
The paper is organized as follows.
Sec.~\ref{sec:pre} provides the necessary background on direct data-driven control and a adversarial attack method against neural networks.
Sec.~\ref{sec:thr} introduces the threat model and presents our proposed perturbation generation algorithms.
Sec.~\ref{sec:alg} presents an efficient gradient computation method via implicit differentiation using the KKT conditions.
Sec.~\ref{sec:def} introduces defense strategies based on regularization and robust data-driven controller design.
Sec.~\ref{sec:exp} presents numerical experiments.
Finally, Sec.~\ref{sec:conc} concludes the paper and discusses future research directions.
Appendix summarizes the basic rules of matrix calculus and exhibits proofs of the propositions.

We denote
the $n$-dimensional identity matrix by $I_n$,
the $n\times m$-dimensional zero matrix by $0_{n,m}$,
the transpose and the Moore-Penrose pseudoinverse of a matrix $M$ by $M^{\sf T}$ and $M^{\dagger}$, respectively,
the trace, the spectral radius, and the minimum singular value of a square matrix $M$ by $\trace(M)$, $\rho(M),$ and $\sigma_{\rm min}(M)$, respectively,
the 2-induced norm, the Frobenius norm, and the element-wise max norm of a matrix $M$ by $\|M\|_2$, $\|M\|_{\rm F}$, and $\|M\|_{\rm max}$, respectively,
the positive and negative (semi)definiteness of a Hermetian matrix $M$ by $M\succ $ ($\succeq$) $0$ and $M\prec$ ($\preceq$) $0$, respectively,
the block diagonal matrix whose diagonal blocks are composed of $M_1$ and $M_2$ by ${\rm diag}(M_1,M_2)$,
the Kronecker product by $\otimes$,
the $(p,q)$ commutation matrix by $C_{p,q}$,
the component-wise sign function by ${\rm sign}(\cdot)$,
the vectorization operator by ${\rm vec}$,
and the Frobenius inner product of two real matrices $(M,N)$ by $\innerproduct{M}{N}\coloneqq{\rm tr}(M^{\sf T}N).$
The subscript for the dimension is often omitted when it is clear from the context.

\section{Preliminaries}
\label{sec:pre}

\subsection{Direct Data-driven LQR Control}
Consider a discrete-time linear time-invariant (LTI) system
$x(t+1)=Ax(t)+Bu(t)$ for $t=0,1,\ldots$,
where $x_t\in\mathbb{R}^n$ is the state and $u_t\in\mathbb{R}^m$ is the control input.
We assume that the pair $(A,B)$ is stabilizable.
We consider the LQR problem~\cite[Chap.~6]{Chen2012Optimal}, which has been widely studied as a benchmark problem.
Specifically, design a static state-feedback control $u_t=Kx_t$ that minimizes the cost function
$\mc{J}(K)=\sum_{i=1}^n\sum_{t=0}^{\infty} \left(x_t^{\sf T}Qx_t + u_t^{\sf T}Ru_t\right)|_{x_0=e_i}$
with $Q\succeq 0$ and $R\succ 0$ where $e_i$ is the $i$th canonical basis vector.
It is known that the cost function can be rewritten as
$\mc{J}(K)=\trace(QP)+\trace(K^{\sf T}RKP)$
where $P\succeq I$ is the controllability Gramian of the closed-loop system when $A+BK$ is Schur, i.e., $\rho(A+BK)<1$.
The objective of direct data-driven control is to design the optimal feedback gain using data of input and state signals without requiring explicit system identification.

The overall implementation of the direct data-driven LQR approach is structured into two distinct phases: an \emph{offline design phase} and an \emph{online operation phase}, which are analogous to the training and test phases commonly seen in machine learning. 
In the offline design phase, historical trajectory data of the system are collected under exploratory input signals.
Using this dataset, the controller is synthesized by solving an optimization problem, which learns the optimal feedback gain.
This design process is entirely performed offline and does not require explicit knowledge of the system matrices $(A,B)$.
Once the controller gain $K$ is computed, the online phase involves real-time operation of the system using the static linear state-feedback law $u_t = Kx_t$.

Here, we review the offline design phase in detail.
It is assumed that the system matrices $(A,B)$ are unknown, but instead, the tuple of $T$-long offline batch data $(Z,X,U)$ that obey the dynamics
$Z=AX+BU,$
where $Z\in\mathbb{R}^{n\times T}$, $X\in\mathbb{R}^{n\times T},$ and $U\in\mathbb{R}^{m\times T}$,
are available.
We denote the collective data by $D\coloneqq[Z^{\sf T}\ X^{\sf T}\ U^{\sf T}]^{\sf T}$.
We assume that
$\iGam\coloneqq [U^{\sf T}\ X^{\sf T}]^{\sf T}$ is full rank,
which is generally necessary for data-driven LQR design~\cite{Van2020Data}.
It is satisfied if the input signal is persistently exciting as shown by the Willems' fundamental lemma~\cite{Willems2005Note}.

The key idea of the approach laid out in~\cite{De2019Formulas} is to parameterize the controller using the available data by introducing a new variable $G\in\mathbb{R}^{T\times n}$ with the relationship
\begin{equation}\label{eq:KI}
[K^{\sf T}\ I]^{\sf T}
=\iGam G.
\end{equation}
Then the closed-loop matrix can be parameterized directly by data matrices as
\begin{equation}\label{eq:ABKXG}
A+BK = [B\ A]\iGam G = ZG.
\end{equation}
The LQR controller design can be formulated as
\begin{equation}\label{eq:prob_ori}
\begin{array}{cl}
\displaystyle{\min_{P,K,G}} & \mc{J}(K)\\
 {\rm s.t.} & ZGPG^{\sf T}Z^{\sf T}-P+I\preceq 0,
 \ P \succeq I,\ \eqref{eq:KI}.
\end{array}
\end{equation}

However, it has been pointed out that the formulation~\eqref{eq:prob_ori} is not robust to disturbance~\cite{Dorfler2022On}.
Specifically, the study considers a control system subject to disturbances given by $x(t+1)=Ax(t)+Bu(t)+d(t)$ with the exogeneous disturbance $d(t)\in\mathbb{R}^n$, which leads to the data relationship
$Z = AX+BU+W$
where $W\in\mathbb{R}^{n\times T}$ is the disturbance matrix.
Then we have $A+BK=(Z-W)G$ instead of~\eqref{eq:ABKXG}, and hence the optimization problem~\eqref{eq:prob_ori} fails to produce the exact solution to the LQR problem.
To address this issue, a regularized formulation has been proposed for enhancing robustness against disturbance:
\begin{equation}\label{eq:prob_reg1}
\begin{array}{cl}
\displaystyle{\min_{P,K,G}} & \mc{J}(K)+\gamma\|\Pi G\|_{\rm F}^2\\
 {\rm s.t.} & ZGPG^{\sf T}Z^{\sf T}-P+I\preceq 0,
 \ P \succeq I,\ \eqref{eq:KI}
\end{array}
\end{equation}
with a constant $\gamma\geq0$ where $\Pi\coloneqq I-\iGam^{\dagger}\iGam$.
The regularizer $\gamma\|\Pi G\|_{\rm F}$ is referred to as certainty-equivalence regularization because it leads to the controller equivalent to the certainty-equivalence indirect data-driven LQR with ordinary least-square estimation of the system model when $\gamma$ is sufficiently large~\cite{Dorfler2022On}.
Note that the reformulated problem can also be converted into the following SDP:
\begin{equation}\label{eq:sdp_form}
\begin{array}{l}
 \min_{L,S} J(L,S,D)\ 
  {\rm s.t.}\ 
F(L,S,D)\succeq 0
\end{array}
\end{equation}
with variables $L\in\mathbb{R}^{T\times n}, S\in\mathbb{R}^{m\times m}$ for given data $D$ where
\begin{equation}
 J\coloneqq {\rm tr}(QXL)+{\rm tr}(S) +\gamma \|\Pi L\|_{\rm F}^2,\ 
 F\coloneqq {\rm diag}(F_1,F_2),
\end{equation}
and
\begin{equation}
 F_1\coloneqq \begin{bmatrix}
      S & VUL\\
      \ast & XL
  \end{bmatrix},\\
 F_2\coloneqq \begin{bmatrix}
      XL-I & ZL\\
      \ast & XL
  \end{bmatrix},
\end{equation}
and $V\succ 0$ is the square root of $R$.
Note that $F(L,S,D)\in\mathbb{R}^{(3n+m)\times (3n+m)}.$
The resulting controller is given by $K=F_K(L,D)$ with
$F_K(L,D)\coloneqq UL(XL)^{-1}$
and $P=XL$ where $(L,S)$ is the optimal solution to~\eqref{eq:sdp_form}.

\subsection{Fast Gradient Sign Method}

The Fast Gradient Sign Method (FGSM) is a widely used technique for efficiently generating adversarial perturbations against trained neural networks~\cite{Goodfellow2015Explaining}.
Given a loss function $\ell(D,Y;\theta)$ of the targeted neural network where $D\in\mathbb{R}^{p\times q}$ is the input data, $Y\in\mathcal{Y}$ is the true label, and $\theta$ is the trained parameter, FGSM aims to craft a perturbation $\Delta\in\mathbb{R}^{p\times q}$ that causes the classifier $f:\mathbb{R}^{p\times q}\to\mathcal{Y}$ to misclassify the input, i.e., $f(D+\Delta)\neq f(X)$.
The perturbation is constrained in magnitude by the element-wise max norm $\|\Delta\|_{\rm max} \leq \epsilon$, where $\epsilon>0$ is a small constant ensuring imperceptibility.

The core idea behind FGSM is to exploit the linear approximation of the loss function with respect to the input:
$\ell(D+\Delta,Y;\theta)\simeq \ell(D,Y;\theta)+ \innerproduct{\nabla_D\ell(D,Y;\theta)}{\Delta}.$
To maximize the loss under the norm constraint, FGSM chooses each entry of $\Delta$ to align with the sign of the gradient:
$\Delta = \epsilon\, {\rm sign}(\nabla_{D}\ell(D,Y;\theta)).$
This perturbation direction maximally increases the loss in a single step under the given constraint.
In practice, FGSM is often used iteratively, gradually increasing $\epsilon$ until misclassification is achieved.

\section{Adversarial Attack Framework}
\label{sec:thr}
\subsection{Threat Model}
\begin{figure}[t]
  \centering
  \includegraphics[width=0.98\linewidth]{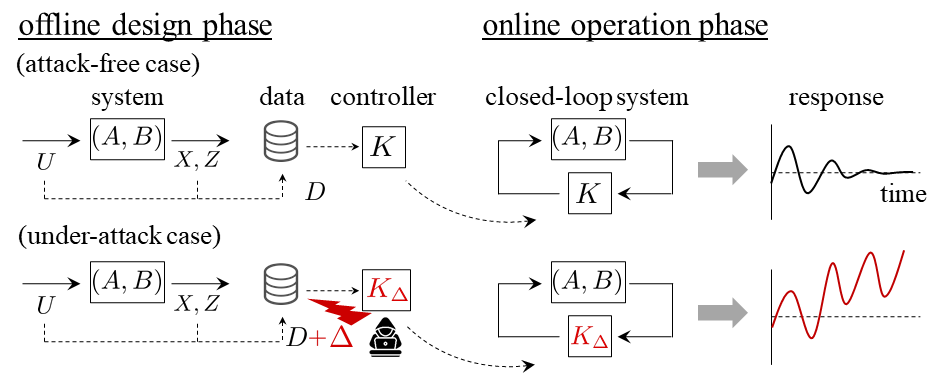}
  \caption{
  Threat model addressed in this study.
  The adversary is capable of manipulating the data $D$ by introducing a perturbation $\Delta$.
  The controller $K_{\Delta}$ designed with the perturbed data $D+\Delta$ may lead to instability of the closed-loop system.
  }
  \label{fig:scenario}
\end{figure}

We consider an adversary whose goal is to destabilize the closed-loop control system by introducing small but malicious perturbations to the input-state data used for controller synthesis.
The adversarial manipulations are designed to render the resulting system unstable, while remaining imperceptible.
Fig.~\ref{fig:scenario} illustrates the overall threat model.

The attack is executed during the offline design phase.
The adversary introduces a perturbation $(\Delta_{Z},\Delta_X,\Delta_U)$ to the clean input and state data $(Z,X,U)$ used in the direct data-driven LQR control algorithm, thereby influencing the resulting controller learned from the perturbed data
\begin{equation}
 (Z_\Delta,X_\Delta,U_\Delta) \coloneqq  (Z+\Delta_Z,X+\Delta_X,U+\Delta_U).
\end{equation}
This type of attack is often referred to as data poisoning attack~\cite{Biggio2012Poisoning}, because it corrupts the training data prior to learning, causing the learned controller to behave in a manner unintended by the designer.
In this context, the ``training data'' corresponds to the historical trajectories used for direct controller synthesis, and the adversary exploits this dependency to inject targeted instabilities.

The adversary can modify the input and state sequences used for control design, but the magnitude of perturbation is bounded.
We constrain the norm of the perturbation matrix $\Delta\coloneqq [\Delta_Z^{\sf T}\ \Delta_X^{\sf T}\ \Delta_U^{\sf T}]^{\sf T}$, i.e., $\|\Delta\|\leq \epsilon$ with a small constant $\epsilon>0$ with a chosen matrix norm $\|\cdot\|$.
Specifically, we adopt the element-wise max norm $\|\cdot\|_{\rm max}$, because it directly limits the largest allowable change to any individual data point.
Alternatively, one may consider a relative perturbation constraint, each of $(\Delta_Z,\Delta_X,\Delta_U)$ is bounded in proportion to the corresponding data matrix, i.e.,
\begin{equation}\label{eq:rel_const}
    \|\Delta_Z\|\leq \epsilon\,\|Z\|,\quad \|\Delta_X\|\leq \epsilon\,\|X\|,\quad \|\Delta_U\|\leq \epsilon\,\|U\|.
\end{equation}
The constraint in~\eqref{eq:rel_const} is interpreted as an attacker-side surrogate for stealthiness: the attacker limits the deviation from the original data to reduce the risk of generating anomalous or implausible observations.
Although the defender does not observe the true dataset, we assume that nominal data are broadly consistent with the defender's prior knowledge of physically plausible ranges or statistical bounds, so that the discrepancy between model-consistent data and the true data is not significant.
Without the prior model used by the defender, it can only control the deviation from the original data as a proxy for reducing exposure risk.
Under these assumptions, proximity to the original data serves as a proxy for remaining within admissible ranges.
Accordingly, the norm constraint should be understood as a practical constraint motivated from the attacker's perspective.
It enforces plausibility without assuming a specific anomaly detection mechanism for the dataset used in the offline controller design.
Note that, in this paper, we consider a simple destabilizing attack objective: once the closed-loop system exhibits a large deviation from the setpoint or anomalous behavior becomes observable, the attacker's goal is achieved.
More sophisticated attack formulations that explicitly model online detection mechanisms are beyond the scope of this work.

We primarily consider a white-box adversary with full knowledge of the system model $(A,B)$, the controller design algorithm, and the clean data $D$.
Although this scenario represents a worst-case upper bound on the potential attack impact, such a highly capable adversary may be unlikely in practice.
To address this limitation, we later extend our analysis to a gray-box setting, where the adversary lacks knowledge of the precise dataset used for controller design.
In this case, a plausible attack strategy is to use hypothetical data $D_{\rm hyp}$ that is consistent with the system dynamics but follows a different trajectory from the training data.
We refer to the effectiveness of such an attack as transferability across data, numerically evaluated in Sec.~\ref{sec:exp}.

\emph{Remark:}
Since system manufacturers often rely on extensive datasets gathered from operational systems or obtained from third-party sources, this type of attack can arise in two ways.
First, an attacker may gain access to the data collection hardware, such as by compromising data loggers, and introduce malicious measurement errors.
For example, prior work has demonstrated physically aware malware capable of intruding programmable logic controllers (PLCs), enabling precise manipulation of collected data at the source~\cite{Garcia2017Hey}.
Second, an attacker could tamper with external data providers or shared repositories supplying the training data.
Supply-chain attacks provide concrete evidence that trusted data or software pipelines can be compromised at scale.
A prominent example is the SolarWinds incident, which illustrates that adversaries can gain white-box access to trusted pipelines prior to deployment~\cite{Martinez2021Software}.
As a result, the system may inadvertently use the corrupted data when tuning its controller.

\subsection{Perturbation Generation Algorithm}

Our goal is to construct an adversarial attack algorithm that is numerically efficient and practically effective, rather than to provide formal optimality guarantees.
We focus on designing a tractable search procedure that performs well in practice, and demonstrate its effectiveness through numerical experiments.

\subsubsection{Directed Gradient Sign Method (DGSM)}
Having established the threat model, we now formulate the perturbation generation task as an optimization problem.
Let $\mc{K}:D \mapsto K$ denote the mapping from the given data $D$ to the controller designed by the direct data-driven control described in Sec.~\ref{sec:pre}.
Then the optimization problem can be formulated as
\begin{equation}
\begin{array}{cl}
    \max_{\Delta} & \rho(A+B\mc{K}(D+\Delta)) \\
    {\rm s.t.} & \|\Delta\|_{\rm max}\leq \epsilon.
\end{array}
\end{equation}

\begin{figure}[t]
  \centering
  \includegraphics[width=0.98\linewidth]{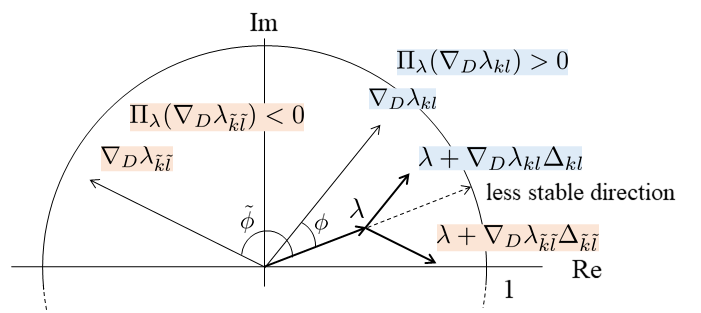}
  \caption{
 Illustration of the function $\Pi_{\lambda}$ in~\eqref{eq:Pi}.
  }
  \label{fig:Pi}
\end{figure}

Our proposed algorithm, Directed Gradient Sign Method (DGSM), is a natural extension of FGSM tailored to the control-theoretic setting of our problem.
A technically crucial difficulty is that the spectral radius is non-differentiable whenever the associated dominant eigenvalue is not simple. To address this issue, we employ a smoothing technique based on the Gelfand approximation~\cite{Kozyakin2009Accuracy}.
Specifically, we replace $\rho(A+BK)$ with the smooth function
\[
 \hat{\rho}(A+BK)\coloneqq \|(A+BK)^{k_{\rm G}}\|^{1/k_{\rm G}}_{\rm F}
\]
with a positive integer $k_{\rm G}$.
Note that $\hat{\rho}(A+BK)\to\rho(A+BK)$ as $k_{\rm G}\to\infty$.
DGSM then selects the perturbation as
\begin{equation}
    \Delta = \epsilon\, {\rm sign}(\nabla_D \hat{\rho}(A+B\mc{K}(D))),
\end{equation}
assuming that the mapping $\mc{K}(D)$, defined through the SDP~\eqref{eq:sdp_form}, is differentialbe.
Geometrically, this perturbation shifts the dominant eigenvalue $\lambda(D+\Delta)$, which corresponds to the spectral radius of the closed-loop matrix, in the direction that reduces stability.
The perturbed dominant eigenvalue is linearly approximated by
\begin{equation}
     \textstyle{\lambda(D+\Delta) \simeq \lambda}+ \innerproduct{\nabla_D\hat{\lambda}}{\Delta},
\end{equation}
where $\nabla_D\hat{\lambda}$ denotes the gradient obtained from the Gelfand approximation.
The perturbation matrix computed by DGSM can be equivalently expressed as
\[
  \Delta_{kl}=\epsilon\, {\rm sign}(\Pi_{\lambda}((\nabla_D\hat{\lambda})_{kl}))
\]
where the subscript $kl$ denotes the $(k,l)$th component and
$\Pi_{\lambda}:\mathbb{C}\to\mathbb{R}$ is an operator defined by
\begin{equation}\label{eq:Pi}
\Pi_{\lambda}(\nabla_D\hat{\lambda}_{kl}):= \Real{\lambda}\Real{\nabla_D\hat{\lambda}_{kl}}+\Imag{\lambda}\Imag{\nabla_D\hat{\lambda}_{kl}}.
\end{equation}
Fig.~\ref{fig:Pi} illustrates a geometric interpretation of the operator $\Pi_{\lambda}$.
Suppose that $\nabla_D\hat{\lambda}_{kl}$ points in the direction of $\lambda$.
More precisely, the angle between $\lambda$ and $\nabla_D\hat{\lambda}_{kl}$, denoted by $\phi$, is less than $\pi/2$, which leads to $\Pi_{\lambda}(\nabla_D\hat{\lambda}_{kl})>0$.
On the other hand, when the angle between $\lambda$ and another element $\nabla_D\hat{\lambda}_{kl}$, denoted by $\tilde{\phi}$, is greater than $\pi/2$, we have $\Pi_{\lambda}(\nabla_D\hat{\lambda}_{\tilde{k}\tilde{l}})<0$.
In both cases, owing to the function $\Pi_{\lambda},$ the perturbed eigenvalue moves closer to the unit circle.
This alignment with the destabilizing direction motivates the name ``Directed'' Gradient Sign Method.
Note that these approximations are introduced to obtain a tractable direction for perturbation design, and their impact is assessed empirically rather than through explicit error analysis.

\subsubsection{Iterative Directed Gradient Sign Method (I-DGSM)}
To enhance the performance of DGSM, we also propose an iterative variant based on DGSM, referred to as Iterative Directed Gradient Sign Method (I-DGSM).
Algorithm~1 provides its algorithm where $\alpha_{\rm step}>0$ is the step size and ${\rm Proj}_{\epsilon}(\hat{\Delta})$ denotes the element-wise truncation function, i.e., its $(k,l)$th component is given by
\begin{equation}
 {\rm Proj}_{\epsilon}(\hat{\Delta})_{kl}=\left\{
 \begin{array}{cl}
 \epsilon\,{\rm sign}(\hat{\Delta}_{kl}) & {\rm if}\ |\hat{\Delta}_{kl}|\geq \epsilon,\\
 \hat{\Delta}_{kl} & {\rm otherwise}. 
 \end{array}
 \right.
\end{equation}
In the relative constraint case with~\eqref{eq:rel_const}, the projection map is individually applied to $(\Delta_Z,\Delta_X,\Delta_U)$ with the corresponding range.
This approach adopts a smoothing projected gradient method~\cite{Zhang2009Smoothing} to more effectively solve the underlying perturbation generation problem.
While global optimality cannot be guaranteed due to the non-convex nature of the bi-level problem, the algorithm employs diminishing step sizes $\alpha_k$ such that $\alpha_k\to 0$ and $\sum_{k=0}^\infty\alpha_k=+\infty$ to stabilize the iterative updates, providing a practical approach to identify severe, stability-compromising perturbations.

\begin{algorithm}[th]
\caption{Iterative Directed Gradient Sign Method (I-DGSM)}
\begin{algorithmic}[1]
\REQUIRE{$\epsilon,\alpha_k,A,B,D,\mathcal{K}$}
\ENSURE{$\Delta$}
\STATE $\Delta_0=0$
\STATE $k \leftarrow 0$
\WHILE{Termination condition not met}
\STATE $\hat{\Delta}_{k+1} \leftarrow \Delta_k + \alpha_k\nabla_D \hat{\rho}(A+B\mathcal{K}(D+\Delta_k))$
\STATE $\Delta_{k+1}\leftarrow {\rm Proj}_{\epsilon}(\hat{\Delta}_{k+1})$
\STATE $k \leftarrow k+1$
\ENDWHILE
\RETURN $\Delta_k$
\end{algorithmic}
\end{algorithm}

In Algorithm 1, defining a theoretically rigorous termination condition is challenging, as it would require prior knowledge of the optimal value.
In practice, we found that simply limiting the maximum number of iterations provides a reliable and stable criterion.
Alternatives, such as monitoring the norm of the smoothed subgradient or tracking the objective function value, can be adopted depending on the system type.

Remark: As noted, the gradient computation relies on the differentiability of the mapping $\mc{K}:D\mapsto K$, where $K$ is obtained by solving the SDP~\eqref{eq:sdp_form}.
The differentiability of solutions to parameterized SDPs is known to be a delicate issue: smoothness of the optimizer generally requires structural regularity conditions such as strict complementarity and second-order sufficient conditions: see e.g.,~\cite{Bonnans2000Perturbation,Bellon2025Parametric}.
Because it is highly challenging to verify the regularity conditions explicitly, in this work, we adopt the standard practical approach of assuming differentiability when computing $\nabla_D \hat{\rho}(A+B\mc{K}(D))$.
Note that, nevertheless, our numerical results in Sec.~\ref{sec:exp} indicate that the proposed gradient-based algorithm performs robustly even without such guarantees.

\section{Efficient Gradient Computation}
\label{sec:alg}

To implement I-DGSM, it is essential to compute the gradient of the spectral radius of the closed-loop matrix with respect to the perturbation matrix.
Assuming differentiability of $L$, the chain rule (see Appendix~\ref{app:mat_der}) leads to the relationship
\begin{equation}\label{eq:deriv_spec_D}
\dfrac{d\hat{\rho}}{dD} = \dfrac{d\hat{\rho}}{d\mc{K}}\left(
\dfrac{\partial F_K}{\partial L} \dfrac{dL}{dD} + \dfrac{\partial F_K}{\partial D}
\right).
\end{equation}
In this equation, the computation of $dL/dD$ is non-trivial because $L$ is the solution to the SDP~\eqref{eq:sdp_form} whose constraints and objective depend on the perturbed data.
A straightforward approach based on numerical differentiation (e.g., central difference method~\cite[Chap.~4]{Burden2015Numerical}) requires solving the SDP $\mathcal{O}((2n+m)T)$ times.
This requirement is computationally expensive and impractical, especially for I-DGSM where the gradient must be computed at every iteration.
This motivates the need for a more efficient and scalable method to compute the required gradients.

\subsection{Implicit Differentiation Approach}
\label{subsec:impdiff}

To address the computational overhead of numerical differentiation, we adopt an \emph{implicit differentiation} approach~\cite{Blondel2022Efficient,Amos2017Optnet,Xu2024Revisiting}, by leveraging the optimality conditions of the optimization problem used in controller synthesis.
Instead of explicitly solving perturbed SDPs repeatedly, we derive an analytical expression for the gradient by differentiating an implicit function defined by $F_{\rm imp}(L(D),D)=0$, where we express the argument $D$ for the optimal solution $L$ to emphasize the dependency.
Applying the chain rule yields
\begin{equation}\label{eq:chain_implicit}
 \dfrac{\partial F_{\rm imp}}{\partial L}\dfrac{dL}{dD}+\dfrac{\partial F_{\rm imp}}{\partial D}=0,
\end{equation}
from which $dL/dD$ can be obtained by solving the linear equation, provided the partial derivatives $\partial F_{\rm imp}/\partial L$ and $\partial F_{\rm imp}/\partial D$ are tractable.

To construct the implicit function, we exploit the KKT conditions for SDP~\eqref{eq:sdp_form}~\cite[Chap.~5]{Boyd2004Convex}
given as
\begin{equation}\label{eq:KKT}
\left\{
\begin{array}{l}
 \partial\mc{L}(L,S,\Lambda,D)/\partial L=0,\\
 \partial\mc{L}(L,S,\Lambda,D)/\partial S=0,\\
 \trace(F(L,S,D)\Lambda^{\sf T})=0,\\
 \Lambda - \Lambda^{\sf T}=0,\\
\end{array}
\right.
\end{equation}
with the primal and dual feasibility conditions
$F(L,S,D)\succeq 0$ and $\Lambda \succeq 0$,
where $\Lambda\in \mathbb{R}^{(3n+m)\times (3n+m)}$ denotes the Lagrange multipliers and the Lagrangian $\mc{L}$ is given by
$\mc{L}(L,S,\Lambda,D) \coloneqq J(L,S,D)-\trace(F(L,S,D)\Lambda^{\sf T})$.
In our approach, we focus on the equality constraints in~\eqref{eq:KKT} by excluding the semidefiniteness constraints, as justified by prior work~\cite{Blondel2022Efficient,Duvenaud2020Deep}.
We hereinafter assume the existence of a continuously differentiable implicit function $(L(D), S(D), \Lambda(D))$ locally around the solution satisfying the KKT conditions, which can be verified by applying the implicit function theorem~\cite[Chap.~7]{Magnus2019Matrix}.
For notational simplicity, we define the four components of the KKT conditions as $G_i=0$ for $i=1,\ldots,4$, where
$G_1\in\mathbb{R}^{1\times nT},\ G_2\in\mathbb{R}^{1\times m^2},\ G_3\in\mathbb{R},\ G_4\in\mathbb{R}^{(3n+m)\times(3n+m)}.$

Differentiating the system~\eqref{eq:KKT} with respect to $D$ yields the linear equation
{\small\begin{equation}\label{eq:lin_SDP}
\begin{aligned}
\underbrace{
\begin{bmatrix}
 \dfrac{\partial G_1}{\partial L} & \dfrac{\partial G_1}{\partial S} & \dfrac{\partial G_1}{\partial \Lambda}\\
 \vdots & \vdots & \vdots\\
 \dfrac{\partial G_4}{\partial L} & \dfrac{\partial G_4}{\partial S} & \dfrac{\partial G_4}{\partial \Lambda}
\end{bmatrix}
}_{\coloneqq H}
\begin{bmatrix}
    \dfrac{\strut dL}{\strut dD}\\
    \dfrac{\strut dS}{\strut dD}\\
    \dfrac{\strut d\Lambda}{\strut dD}
\end{bmatrix}
+
\begin{bmatrix}
    \dfrac{\partial G_1}{\partial D}\\
    \vdots\\
    \dfrac{\partial G_4}{\partial D}
\end{bmatrix}=0,
\end{aligned}
\end{equation}
}which corresponds to~\eqref{eq:chain_implicit}.
Note that the coefficient matrices are comparatively easy to compute numerically because we have closed-form expressions for $\mc{L},G_3,G_4$.
\subsection{Rank Complement}
\label{subsec:rankcomp}

The linear system~\eqref{eq:lin_SDP}, derived from differentiating the KKT conditions, can be written compactly using a coefficient matrix $H\in\mathbb{R}^{(nT+m^2+1+(3n+m)^2)\times(nT+m^2+(3n+m)^2)}$.
However, this equation is generally underdetermined due to rank deficiency of $H$:
\begin{prop}\label{prop:rank-deficient}
 The matrix $H$ has a non-trivial kernel for any $L,S,\Lambda,D$.
\end{prop}

This rank deficiency arises from the structure of the KKT conditions, in particular the scalar complementarity condition $\trace(F\Lambda^{\sf T}) = 0$.
While this constraint is sufficient to certify optimality, it introduces only a single scalar equation, which limits the rank of $H$ and prevents unique identification of the gradient $dL/dD$.
To resolve this issue, we seek to inject additional valid constraints that are consistent with optimality but provide more structural information.
To this end, we leverage the following result:
\begin{prop}\label{prop:trace_reduction}
 For any positive semidefinite matrices $S_1$ and $S_2$, $\trace(S_1S_2)=0$ if and only if $S_1S_2=0$.
\end{prop}

Proposition~\ref{prop:trace_reduction} implies that, under the positive semidefinite assumption, the scalar condition $G_3=\trace(F\Lambda^{\sf T}) = 0$ can be equivalently replaced by the matrix equality $\bar{G}_3 \coloneqq F\Lambda^{\sf T} = 0$.
This matrix equality introduces multiple scalar equations and thus contributes more rank to the system.
By incorporating this into the KKT structure, we obtain the following augmented system:
{\small
\begin{equation}\label{eq:lin_SDP_aug}
\begin{aligned}
\begin{bmatrix}
 \dfrac{\stp G_1}{\stp L} & \dfrac{\stp G_1}{\stp S} & \dfrac{\stp G_1}{\stp \Lambda}\\
 \dfrac{\stp G_2}{\stp L} & \dfrac{\stp G_2}{\stp S} & \dfrac{\stp G_2}{\stp \Lambda}\\
 \dfrac{\stp \bar{G}_3}{\stp L} & \dfrac{\stp \bar{G}_3}{\stp S} & \dfrac{\stp \bar{G}_3}{\stp \Lambda}\\
 \dfrac{\stp G_4}{\stp L} & \dfrac{\stp G_4}{\stp S} & \dfrac{\stp G_4}{\stp \Lambda}
\end{bmatrix}
\begin{bmatrix}
    \dfrac{\strut dL}{\strut dD}\\
    \dfrac{\strut dS}{\strut dD}\\
    \dfrac{\strut d\Lambda}{\strut dD}
\end{bmatrix}
+
\begin{bmatrix}
    \dfrac{\stp G_1}{\stp D}\\
    \dfrac{\stp G_2}{\stp D}\\
    \dfrac{\stp \bar{G}_3}{\stp D}\\
    \dfrac{\stp G_4}{\stp D}
\end{bmatrix}=0.
\end{aligned}
\end{equation}}
The complete gradient computation procedure is summarized in Algorithm~2.
Numerical examples in Sec.~\ref{sec:exp} empirically indicate the uniqueness of the solution to the augmented system~\eqref{eq:lin_SDP_aug}.

\begin{algorithm}[th]
\caption{Gradient Computation using Implicit Differentiation}
\begin{algorithmic}[1]
\REQUIRE{$J,F,D$}
\ENSURE{$d L/d D$}
\STATE Solve~\eqref{eq:sdp_form} given $D$ and obtain optimal solutions $(L,S,\Lambda)$
\STATE Compute the coefficient matrices in~\eqref{eq:lin_SDP_aug} around $(L,S,\Lambda)$
\STATE Solve~\eqref{eq:lin_SDP_aug} and obtain $(dL/dD,dS/dD,d\Lambda/dD)$
\RETURN $d L/ d D$
\end{algorithmic}
\end{algorithm}

\subsection{Further Efficient Computation}
\label{subsec:further_efficient}

Computing the coefficient matrices in~\eqref{eq:lin_SDP_aug} via numerical differentiation requires evaluating $\mc{L}$ many times.
This leads to a significant computational burden and can become a bottleneck for calculation precision.
To address this, we derive closed-form expressions for the required derivatives.
We begin by presenting the analytical derivatives used in~\eqref{eq:deriv_spec_D}.
\begin{prop}\label{prop:der_FK}
We have
\begin{equation}
\begin{aligned}
 & d\hat{\rho}/dK = k_{\rm G}^{-1}\|A_{\rm cl}^{k_{\rm G}}\|_{\rm F}^{1/k_{\rm G}-2}{\rm vec}(A_{\rm cl}^{k_{\rm G}})^{\sf T}\\
 & \quad \quad \quad \quad \quad \quad \textstyle{\cdot(\sum_{j=0}^{k_{\rm G}-1} A_{\rm cl}^{k_{\rm G}-1-j}\otimes A_{\rm cl}^{j{\sf T}}),}\\
 & \partial F_K/\partial L = (XL)^{-{\sf T}}\otimes(U-F_KX),\\
 & \partial F_K/\partial D = ((XL)^{-{\sf T}}L^{\sf T})\otimes (E_U-F_KE_X)
\end{aligned}
\end{equation}
where $A_{\rm cl}\coloneqq A+BK$, $E_U\coloneqq[0_{m,n}\ 0_{m,n}\ I_m]$, and $E_X\coloneqq[0_{n,n}\ I_n\ 0_{n,m}]$.
\end{prop}

\setlength{\arraycolsep}{3pt}
Subsequently, we derive the derivative of the Lagrangian.
\begin{prop}\label{prop:der_Lag}
We have
\begin{equation}\label{eq:LLLS}
\begin{aligned}
& \dfrac{\partial\mc{L}}{\partial{L}} = \vecT{X^{\sf T}Q^{\sf T}}+2\gamma \vecT{\Pi L} - \vecT{\Lambda} \dfrac{\partial F}{\partial L},\\
& \dfrac{\partial\mc{L}}{\partial{S}} = \vecT{I_m}-\vecT{\Lambda} \dfrac{\partial F}{\partial S},
\end{aligned}
\end{equation}
with
\begin{equation}\label{eq:FF1F2}
\begin{aligned}
& \partial F/\partial L = E_1 \partial F_1/\partial L + E_2 \partial F_2/\partial L,\\
& \partial F/\partial S = E_1 \partial F_1/\partial S + E_2 \partial F_2/\partial S,\\
\end{aligned}
\end{equation}
where $E_1\coloneqq \bar{E}_1\otimes \bar{E}_1$, $E_2\coloneqq \bar{E}_2\otimes \bar{E}_2$,
\begin{equation}
\begin{aligned}
& \bar{E}_1\coloneqq\begin{bmatrix}
    I_{n+m}\\\ 0_{2n,n+m}
\end{bmatrix},\quad
\bar{E}_2 \coloneqq \begin{bmatrix}
    0_{n+m,2n}\\\ I_{2n}
\end{bmatrix},
\end{aligned}
\end{equation}
and the derivatives of $F_1$ and $F_2$ are given in~\eqref{eq:der_Lag}.
\newcounter{MYtempeqncnt}
\begin{figure*}
\setcounter{MYtempeqncnt}{\value{equation}}
\setcounter{equation}{\value{equation}}
{\small\begin{equation}\label{eq:der_Lag}
\begin{aligned}
& \frac{\partial F_1}{\partial L} = \begin{bmatrix}
    0_{m,n}\\ I_n
\end{bmatrix}\otimes\begin{bmatrix}
    VU\\0_{n,T}
\end{bmatrix}+\left(
\begin{bmatrix}
    VU\\0_{n,T}
\end{bmatrix}\otimes\begin{bmatrix}
    0_{m,n}\\I_n
\end{bmatrix}\right)C_{T,n}+
\begin{bmatrix}
    0_{m,n}\\I_n
\end{bmatrix}\otimes
\begin{bmatrix}
    0_{m,T}\\X
\end{bmatrix},\\
& \frac{\partial F_2}{\partial L} =
\begin{bmatrix}
    I_n \\ 0_{n,n}
\end{bmatrix} \otimes
\begin{bmatrix}
    X \\ 0_{n,T}
\end{bmatrix} +
\begin{bmatrix}
    0_{n,n}\\ I_n
\end{bmatrix} \otimes
\begin{bmatrix}
    Z\\ 0_{n,T}
\end{bmatrix} + \left(
 \begin{bmatrix}
     Z\\ 0_{n,T}
 \end{bmatrix}\otimes
 \begin{bmatrix}
     0_{n,n}\\ I_n
 \end{bmatrix}
 \right)C_{T,n} +
 \begin{bmatrix}
     0_{n,n}\\ I_n
 \end{bmatrix}\otimes
 \begin{bmatrix}
     0_{n,T}\\ X
 \end{bmatrix},\\
& \frac{\partial F_1}{\partial S} = \begin{bmatrix}
    I_m\\0_{n,m}
\end{bmatrix}\otimes
\begin{bmatrix}
    I_m\\0_{n,m}
\end{bmatrix},\ \frac{\partial F_2}{\partial S} = 0_{4n^2,m^2}.
\end{aligned}
\end{equation}}
\setcounter{equation}{\value{MYtempeqncnt}+1}
\hrulefill
\end{figure*}
\end{prop}

Finally, we obtain the coefficient matrices in~\eqref{eq:lin_SDP_aug}.
\begin{prop}\label{prop:ana_form_coe_mat}
We have
{\small\begin{equation}
\begin{aligned}
& \frac{\partial G_1}{\partial L} = 2\gamma (I_n\otimes \Pi),\ 
\frac{\partial G_1}{\partial S} = 0_{nT,m^2},\ \frac{\partial G_1}{\partial \Lambda} = -\frac{\partial F}{\partial L}^{\sf T},\\
& 
\begin{split}
\frac{\partial G_1}{\partial D} & = C_{n,T}(I_T\otimes(QE_X))+2\gamma(L^{\sf T}\otimes I_T)\frac{d\Pi}{dD}\\
&\quad -(I_{nT}\otimes\vecT{\Lambda})\frac{\partial^2F}{\partial D\partial L},
\end{split}\\
 & \frac{\partial G_2}{\partial L} = 0_{m^2,nT},\ 
\frac{\partial G_2}{\partial S} = 0_{m^2,m^2},\ \frac{\partial G_2}{\partial \Lambda} = -\frac{\partial F}{\partial S}^{\sf T},\\
& \frac{\partial G_2}{\partial D}=-(I_{m^2}\otimes \vecT{\Lambda})\frac{\partial^2 F}{\partial D\partial S},\\
& \frac{\partial G_3}{\partial L} = (\Lambda\otimes I_{\hat{n}})\frac{\partial F}{\partial L},\
\frac{\partial G_3}{\partial S} = (\Lambda\otimes I_{\hat{n}})\frac{\partial F}{\partial S},\\
\end{aligned}
\end{equation}
\begin{equation}
\begin{aligned}
&\frac{\partial G_3}{\partial \Lambda} = (I_{\hat{n}}\otimes F)C_{\hat{n},\hat{n}},\ 
\frac{\partial G_3}{\partial D} = (\Lambda\otimes I_{\hat{n}})\frac{\partial F}{\partial D},\\
& \frac{\partial G_4}{\partial L}=0_{\hat{n}^2,nT},\ \frac{\partial G_4}{\partial S}=0_{\hat{n}^2,m^2},\\
& \frac{\partial G_4}{\partial \Lambda}=I_{\hat{n}^2}-C_{\hat{n},\hat{n}},\ \frac{\partial G_4}{\partial D}=0_{\hat{n}^2,\bar{n}T},
\end{aligned}
\end{equation}}
where the derivatives are given by
{\small\begin{equation}
\begin{aligned}
& \frac{\partial F}{\partial D} = E_1 \frac{\partial F_1}{\partial D} + E_2 \frac{\partial F_2}{\partial D},\\
& \frac{\partial^2 F}{\partial D \partial L} = (I_{nT}\otimes E_1) \frac{\partial^2F_1}{\partial D \partial L} + (I_{nT}\otimes E_2)
\frac{\partial^2F_2}{\partial D \partial L},\\
&\frac{d\Pi}{dD} = \frac{\partial \Pi}{\partial X}(I_T\otimes E_X)+\frac{\partial \Pi}{\partial U}(I_T\otimes E_U),\\
&\frac{d\Pi}{dX} = -(\iGam \otimes I_T)\frac{d \iGam^\dagger}{d \iGam} \frac{\partial \iGam}{\partial X}
-(I_T\otimes \iGam^\dagger)
\frac{\partial \iGam}{\partial X}
,\\
\end{aligned}
\end{equation}
\begin{equation}
\begin{aligned}
&\frac{d\Pi}{dU} = -(\iGam \otimes I_T)\frac{d \iGam^\dagger}{d \iGam}\frac{\partial \iGam}{\partial U}-(I_T\otimes \iGam^\dagger)\frac{\partial \iGam}{\partial U},\\
& \frac{\partial \iGam}{\partial X} = I_T\otimes
\begin{bmatrix}
    0_{m,n}\\I_n
\end{bmatrix},\quad
\frac{\partial \iGam}{\partial U} = I_T\otimes \begin{bmatrix}
    I_m\\0_{n,m}
\end{bmatrix}
\end{aligned}
\end{equation}
with~\eqref{eq:der_Lag},~\eqref{eq:der_D},~$\hat{n}\coloneqq 3n+m$,~$\bar{n}\coloneqq 2n+m$, and
\begin{equation}
\begin{aligned}
&C_1 \coloneqq I_n\otimes C_{T,m+n} \otimes I_{m+n},\quad
C_2 \coloneqq I_n\otimes C_{T,2n} \otimes I_{2n},\\
& \tilde{C}_1 \coloneqq I_T\otimes C_{n,m+n}\otimes I_{m+n},\quad
\tilde{C}_2 \coloneqq I_T\otimes C_{n,2n}\otimes I_{2n}.
\end{aligned}
\end{equation}}
\newcounter{MYtempeqncnt2}
\begin{figure*}
\setcounter{MYtempeqncnt2}{\value{equation}}
\setcounter{equation}{\value{equation}}
{\small\begin{equation}\label{eq:der_D}
\begin{aligned}
& \frac{\partial F_1}{\partial D}=
\begin{bmatrix}
    0_{m,T}\\ L^{\sf T}
\end{bmatrix}\otimes
\begin{bmatrix}
    VE_U\\0_{n,\bar{n}}
\end{bmatrix}+
\left(
\begin{bmatrix}
    VE_U\\ 0_{n,\bar{n}}
\end{bmatrix}\otimes
\begin{bmatrix}
    0_{m,T}\\L^{\sf T}
\end{bmatrix}
\right)C_{\bar{n},T}+
\begin{bmatrix}
    0_{m,T}\\L^{\sf T}
\end{bmatrix}\otimes
\begin{bmatrix}
    0_{m,\bar{n}}\\E_X
\end{bmatrix},\\
& \frac{\partial F_2}{\partial D} = \begin{bmatrix}
    L^{\sf T}\\0_{n,T}
\end{bmatrix}\otimes
\begin{bmatrix}
    E_X\\0_{n,\bar{n}}
\end{bmatrix}+
\begin{bmatrix}
    0_{n,T}\\L^{\sf T}
\end{bmatrix}\otimes
\begin{bmatrix}
    E_Z\\0_{n,\bar{n}}
\end{bmatrix}+
\left(
\begin{bmatrix}
    E_Z\\0_{n,\bar{n}}
\end{bmatrix}\otimes
\begin{bmatrix}
    0_{n,T}\\L^{\sf T}
\end{bmatrix}
\right)C_{\bar{n},T}+
\begin{bmatrix}
    0_{n,T}\\L^{\sf T}
\end{bmatrix}\otimes
\begin{bmatrix}
    0_{n,\bar{n}}\\E_X
\end{bmatrix},\\
& 
\begin{split}
\frac{\partial^2 F_1}{\partial D\partial L} &= C_1\left(\vecf{
\begin{bmatrix}
0_{m,n}\\I_n    
\end{bmatrix}}\otimes I_{(m+n)T} \right)
\left(
I_T\otimes
\begin{bmatrix}
    VE_U\\ 0_{n,\bar{n}}
\end{bmatrix}
\right)
+
C_1\left(\vecf{
\begin{bmatrix}
0_{m,n}\\I_n    
\end{bmatrix}}\otimes I_{(m+n)T} \right)
\left(
I_T\otimes
\begin{bmatrix}
    0_{m,\bar{n}}\\E_X
\end{bmatrix}
\right)\\
&\quad +(C_{n,T}^{\sf T}\otimes I_{(m+n)^2})\tilde{C}_1\left(
I_{(m+n)T}\otimes \vecf{\begin{bmatrix}
    0_{m,n}\\I_n
\end{bmatrix}}
\right)
\left(
I_T\otimes\begin{bmatrix}
    VE_U\\0_{n,\bar{n}}
\end{bmatrix}
\right),
\end{split}\\
&\begin{split}
\frac{\partial^2 F_2}{\partial D \partial L} & = C_2
\left(
\vecf{\begin{bmatrix}
    I_n\\0_{n,n}
\end{bmatrix}}\otimes I_{2nT}
\right)
\left(
I_T\otimes\begin{bmatrix}
    E_X\\ 0_{n,\bar{n}}
\end{bmatrix}
\right) + C_2
\left(
\vecf{\begin{bmatrix}
    0_{n,n}\\I_n
\end{bmatrix}}\otimes I_{2nT}
\right)
\left(
I_T\otimes\begin{bmatrix}
    E_Z\\ 0_{n,\bar{n}}
\end{bmatrix}
\right)\\
 &\quad + \left(
 C_{n,T}^{\sf T}\otimes I_{4n^2}
 \right)\tilde{C}_2
\left(
I_{2nT}\otimes
\vecf{\begin{bmatrix}
    0_{n,n}\\I_n
\end{bmatrix}}
\right)
\left(
I_T\otimes\begin{bmatrix}
    E_Z\\ 0_{n,\bar{n}}
\end{bmatrix}
\right)
+ C_2
\left(
 \vecf{\begin{bmatrix}
     0_{n,n}\\ I_n
 \end{bmatrix}}\otimes
 I_{2nT}
\right)
\left(
I_T\otimes\begin{bmatrix}
    0_{n,\bar{n}}\\E_X
\end{bmatrix}
\right),
\end{split}\\
&\frac{d\iGam^\dagger}{d\iGam} = ((\iGam \iGam^{\sf T})^{-{\sf T}} \otimes I_T)C_{n+m,T}
 -(I_{m+n}\otimes \iGam^{\sf T})( (\iGam \iGam^{\sf T})^{-{\sf T}} \otimes (\iGam \iGam^{\sf T})^{-1})
 ( (\iGam \otimes I_{m+n}) + (I_{m+n}\otimes \iGam))C_{m+n,T}.
\end{aligned}
\end{equation}}
\setcounter{equation}{\value{MYtempeqncnt2}+1}
\hrulefill
\end{figure*}
\end{prop}
\setlength{\arraycolsep}{5pt}

This closed-form computation of derivatives is significantly more efficient and accurate than numerical differentiation as observed in the numerical experiments.

\subsection{Memory-efficient Implementation}

While leveraging analytic second-derivative formulas provides computational advantages, it often incurs substantial memory overhead.
For instance, the second derivative of the matrix-valued function $F\in\mathbb{R}^{(3n+m)\times(3n+m)}$ with respect to $L\in\mathbb{R}^{T\times n}$ and $D\in\mathbb{R}^{(2n+m)\times T}$ can result in a dense matrix of size $(3n+m)^2nT\times(2n+m)T$.
For modest problem sizes, such as $n=m=T=10$, the number of elements in this matrix reaches 48,000,000.
Storing and manipulating such large matrices can pose serious memory and scalability challenges.

To resolve this issue, we propose a memory-efficient implementation strategy based on an adjoint-based method~\cite{Bell2008Algorithmic} supported by automatic differentiation (AD)~\cite{Paszke2017Automatic} for modest-scale problems.
Specifically, we explicitly utilize the matrix expressions of $\partial \mc{L}/\partial L=G_1$ and $\partial \mc{L}/\partial S=G_2$ as well as the matrices $G_i$ for $i=3,4,5,6$, derived in the previous subsection.
Now consider computing
\[
 \dfrac{d\hat{\rho}}{d\mc{K}}\dfrac{\partial F_K}{\partial L}\dfrac{dL}{dD} = \left[\dfrac{d\hat{\rho}}{d\mc{K}}\quad 0_{1\times m^2}\quad 0_{1\times(3n+m)^2}\right] \left[
 \begin{array}{c}
 dL/dD\\
 dS/dD\\
 d\Lambda/dD
 \end{array}
 \right].
\]
Using the implicit-differentiation identity~\eqref{eq:lin_SDP_aug}, we can apply the adjoint framework (see Appendix~\ref{app:mem_eff} for technical details).
In this setup, no second-order derivative matrix is ever materialized; all higher-order information is accessed only through Jacobian–vector or vector–Jacobian products.
This matrix-free adjoint approach effectively removes the memory bottleneck while still enabling the efficient sensitivity computations required by the proposed algorithm.

\section{Defense Strategies}
\label{sec:def}

Countermeasures for adversarial attacks can be classified into two categories: reactive and proactive strategies~\cite{Yuan2019Adversarial}.
While reactive strategies mainly include detection of attacks, proactive strategies include robustness enhancement of the designed architecture.
In this section, we propose two proactive defense methods that enhance closed-loop stability.

\subsection{Stability Enhancement by Regularization}
We expect that the regularization introduced in~\eqref{eq:prob_reg1} can enhance stability against adversarial attacks, similar to its effect on disturbance rejection.
Although its effectiveness in the context of disturbances has been theoretically analyzed in~\cite{Dorfler2022On}, this analysis does not directly extend to adversarial perturbations due to fundamental differences between them.
For example, while adversarial attacks are added to the data afterward, disturbances enter the internal loop and affect all subsequent outputs.
Additionally, adversarial attacks target input data, while disturbances do not alter the inputs themselves.
These distinctions necessitate a separate analysis to understand how the proposed regularization contributes to stability under adversarial conditions.

To quantify the relative strength of the nominal data versus adversarial perturbations, we introduce a signal-to-perturbation ratio, which measures how large the unperturbed data matrices are compared to their adversarial corruptions.
This quantity plays a role similar to a robustness margin: when the perturbation is sufficiently small relative to the nominal signal, the controller synthesized from the corrupted data remains stabilizing.
Formally, let $\mu$ denote the signal-to-perturbation ratio defined as
\begin{equation}
    \mu \coloneqq
    {\rm min}
    \left(
    \|Z\|_2/\|\Delta_Z\|_2,
    \sigma_{\rm min}(\iGam)/\|\Delta_{\iGam}\|_2
    \right)
\end{equation}
where $\Delta_{\iGam}\coloneqq [\Delta_U^{\sf T}\ \Delta_X^{\sf T}]^{\sf T}$.
A larger value of $\mu$ indicates that the perturbation is small relative to the available data, suggesting that stability should be preserved.

The following proposition makes this intuition precise by providing a sufficient condition on $mu$ under which the closed-loop system remains stable.
\begin{prop}\label{prop:defense_regularization}
The closed-loop system with the controller designed through~\eqref{eq:prob_reg1} with perturbed data $(Z_\Delta,X_\Delta,U_\Delta)$ is stable if
\begin{equation}
    \mu > \left(-1+\sqrt{1+\dfrac{1}{2\|Z\|_2^2\|\MDel\|_2}}\right)^{-1}
\end{equation}
where $\MDel\coloneqq \GDel\PDel\GDel^{\sf T}$ and $(\PDel,\KDel,\GDel)$ denotes the optimal solution to~\eqref{eq:prob_reg1} under adversarial perturbation $\Delta$.
\end{prop}

Proposition~\ref{prop:defense_regularization} establishes that a high signal-to-perturbation ratio guarantees stability under adversarial conditions.
The regularization term $\|\Pi G\|$ in~\eqref{eq:prob_reg1} serves to suppress the magnitude of the variables $G$ and $P$, which in turn reduces the spectral norm $\|M_{\Delta}\|_2$ and thereby enlarges the range of admissible values for the signal-to-perturbation ratio $\mu$.
Furthermore, the proposition highlights the importance of selecting data with large $\sigma_{\min}(\iGam)$ and $\|Z\|_2$, as such data enhance the robustness of the resulting controller against adversarial perturbations.
However, we note that Proposition~\ref{prop:defense_regularization} serves as an explanatory justification for the effectiveness of regularization, rather than a criterion to be checked in practice, since the bound depends on the attacked solution, which is not known a priori.

\setlength{\arraycolsep}{3pt}
\subsection{Stability Guarantee by Robust Data-driven Control}
We propose an alternative defense strategy based on robust data-driven control~\cite{Kaminaga2025DataACC,Kaminaga2025Data}, which can provide formal guarantees against a prescribed class of data perturbations.
Our key insight is that the clean data lies within a neighborhood of the collected data, even if the latter is slightly perturbed.
By designing a controller that guarantees closed-loop stability for all systems consistent with such perturbations, the resulting controller becomes insensitive to data perturbations.

Let the set of admissible perturbations be defined as $\mc{D}\coloneqq \{\Delta:\|\Delta\|_{\rm max}\leq \epsilon\}$.
We introduce the corresponding set of systems consistent with data as $\Sigma\coloneqq\{(A,B):\exists\Delta\in\mc{D}\ {\rm s.t.}\ Z_\Delta = AX_\Delta+BU_\Delta\}$.
We enforce quadratic stability by requiring the existence of a positive definite matrix $P$ such that
\[
 P-(A+BK)P(A+BK)^{\sf T} \succ 0,\quad \forall(A,B)\in\Sigma.
\]
Existing robust data-driven control techniques allow this condition to be reformulated as tractable linear matrix inequalities (LMIs); see~\cite{Kaminaga2025DataACC,Kaminaga2025Data} for details.
By additionally enforcing optimality with respect to quadratic performance for the nominal data, we arrive at the following SDP:
\begin{equation}\label{eq:SDP_robust}
\begin{array}{cl}
\displaystyle{{\rm Find}} & L,S,\alpha\geq 0,\beta\geq 1\\
{\rm s.t.} & \trace(QX_\Delta L)+\trace(S)\leq J_{\rm LQR},\ F(L,S,D_\Delta)\succeq 0,\\
 & {\small \begin{bmatrix}
     X_\Delta L-\beta I_n & 0_{n,n} & 0_{n,m} & 0_{n,n}\\
     \ast & -X_\Delta L &  -L^{\sf T}U_\Delta^{\sf T} & 0_{n,n}\\
     \ast & \ast & 0_{m,m} & U_\Delta L\\
     \ast & \ast & \ast & X_\Delta L
 \end{bmatrix}-\alpha N_\Delta \succeq 0,}
\end{array}
\end{equation}
with $J_{\rm LQR}>0$,
\begin{equation}
   N_\Delta\coloneqq
   {\rm diag}\left(
    \bar{D}_\Delta^{\sf T}
    \begin{bmatrix}
        \epsilon^2 (2n+m) T I_{2n+m} & 0\\
        0 & -I_T
    \end{bmatrix}
    \bar{D}_\Delta,0_{n,n}\right)
\end{equation}
and $\bar{D}_\Delta\coloneqq [I_{2n+m}\ [Z_\Delta^{\sf T}\ -X_\Delta^{\sf T}\ -U_\Delta^{\sf T}]^{\sf T}]^{\sf T}$.
The resulting controller is given by $K=F_K(L,D_\Delta)$.
Its properties can be summarized by the following proposition.
\begin{prop}\label{prop:defense_robust}
Consider the controller $K$ designed via~\eqref{eq:SDP_robust}.  
In the absence of perturbations (i.e., when $\Delta = 0$), the resulting controller achieves the nominal performance $\mc{J}(K)\leq J_{\rm LQR}$.  
Furthermore, in the presence of perturbations, the closed-loop system remains stable for any $\Delta$ satisfying $\|\Delta\|_{\max} \leq \epsilon$.
\end{prop}

\setlength{\arraycolsep}{5pt}

In contrast to the regularization-based defense strategy, the robust data-driven control approach offers a key advantage in that it provides a formal performance and stability guarantee.
In addition, Proposition~\ref{prop:defense_robust} guarantees that the proposed controller not only recovers optimal performance in the nominal setting but also ensures robust stability under bounded perturbations, thereby bridging optimality and robustness within a unified design framework.
Further, in the relative constraint case with~\eqref{eq:rel_const}, $\epsilon$ in $N_{\Delta}$ is replaced with $\epsilon\,\max(\|D\|_{\rm max})$.
Note that feasibility of the proposed SDP depends on the size of the perturbation bound $\epsilon$.
When $\epsilon$ is sufficiently small, the set of systems consistent with the data remains close to the nominal one, and the problem is feasible as long as the performance requirement $J_{\rm LQR}$ is mild.
As $\epsilon$ increases, the corresponding uncertainty set is enlarged, making the robust stability constraints increasingly restrictive and potentially leading to infeasibility.

\section{Experimental Evaluation}
\label{sec:exp}

\subsection{Experimental Setup}
\emph{Benchmarks:}
To validate the generality and effectiveness of the proposed methods, we conduct numerical experiments on several benchmark LTI systems, including five of standard control models provided by the University of Michigan Control Tutorials for MATLAB and Simulink (CTMS)~\cite{William1999Control,Dawn2025Tilbury}, widely used in both education and research.
Specifically, we consider the following systems: motor position control (MP), suspension system (SS), inverted pendulum (IP), aircraft pitch control (AP), and ball and beam (BB).
These systems collectively exhibit a range of dynamic behaviors including instability, underactuation, and lightly damped oscillations.
In addition, we include a linearized model of a triple tank system (TT)~\cite{Blanke2016Diagnosis,Milosevic2019Estimating}, which represents a prototypical process control scenario involving interconnected fluid dynamics.
The inclusion of TT complements the CTMS models and ensures that the proposed approach is applicable to a diverse array of LTI systems across multiple engineering domains.
Although each system has a relatively small state-space dimension (fewer than five states), they serve as fundamental components in larger applications.
Investigating their vulnerabilities therefore provides important insights into the reliability of data-driven control in broader real-world scenarios.

\emph{Baseline Method:} 
To provide a baseline for comparison, we consider a \emph{simple random attack strategy} that perturbs the data by sampling uniformly within the feasible perturbation set.
Specifically, the perturbation matrix $\Delta$ is generated such that each entry is independently drawn from a uniform distribution over an interval that ensures $\|\Delta\|_{\rm max} \leq \epsilon$.
This method does not exploit system structure or optimization but serves as a natural reference to assess the impact of more sophisticated, adversarially optimized attacks.
By comparing against this baseline, we highlight the relative severity and effectiveness of targeted perturbations in degrading stability.

\emph{Attack Impact Metric:}
To quantitatively assess the effectiveness of adversarial perturbations, we adopt the \emph{attack success rate (ASR)} as the primary metric.
This ASR is defined as the ratio of the number of instances in which the controller designed from adversarially perturbed data results in an unstable closed-loop system to the total number of tested instances.
This metric highlights how often an attack successfully destabilizes a previously stable system and serves as a direct indicator of the relative strength of different perturbation methods.

\emph{Parameters:}
All the systems are converted to discrete-time systems using an ideal sampler and a zero-order hold with the sampling period $T_{\rm s}=0.1 s$.
The time horizon is set to $T=2(n+m)$.
The input signal and the initial state are randomly and independently generated by the normal distribution, i.e., $u_t \sim \mc{N}(0,I_m)$ for $t=0,\ldots,T-1$ and $x_0\sim\mc{N}(0,I_n)$.
The state trajectory $x_t$ for $t=1,\ldots,T$ is generated by the dynamics.
The LQR weight matrices are set to $Q=I_n$ and $R=0.01I_m$.
The step size at each iteration of I-DGSM is set to $\alpha_{\rm step}=\epsilon\,\|D\|_{\rm max}$.
The termination condition of I-DGSM is met when either $\rho(A + B\mc{K}(D_k)) \geq 1$ or the iteration count exceeds 50, i.e., $k > 50$.
The relative perturbation constraint~\eqref{eq:rel_const} is adopted in all instances.
Note that the persistence of excitation property holds in all instances.
All reported averages are computed over different data realizations.

\emph{Computational Environment:}
All numerical experiments were performed on a workstation with 3.10 GHz and 12 cores CPU and 64 GB RAM.
Computations were implemented in MATLAB\textsuperscript{\textregistered} (version R2024b)~\cite{Matlab2024} with Python 3.12.10, using PyTorch 2.9.1 for AD~\cite{Paszke2017Automatic}.
The adjoint equation was solved using the Least Squares Minimal Residual (LSMR) method implemented with the LinearOperator interface in SciPy 1.16.3.
The source code used in the numerical experiments is available online~\cite{Sasahara2025Adversraial_GitHub}.

\subsection{Visualization}

\begin{figure*}[t]
  \centering
  \includegraphics[width=0.98\linewidth]{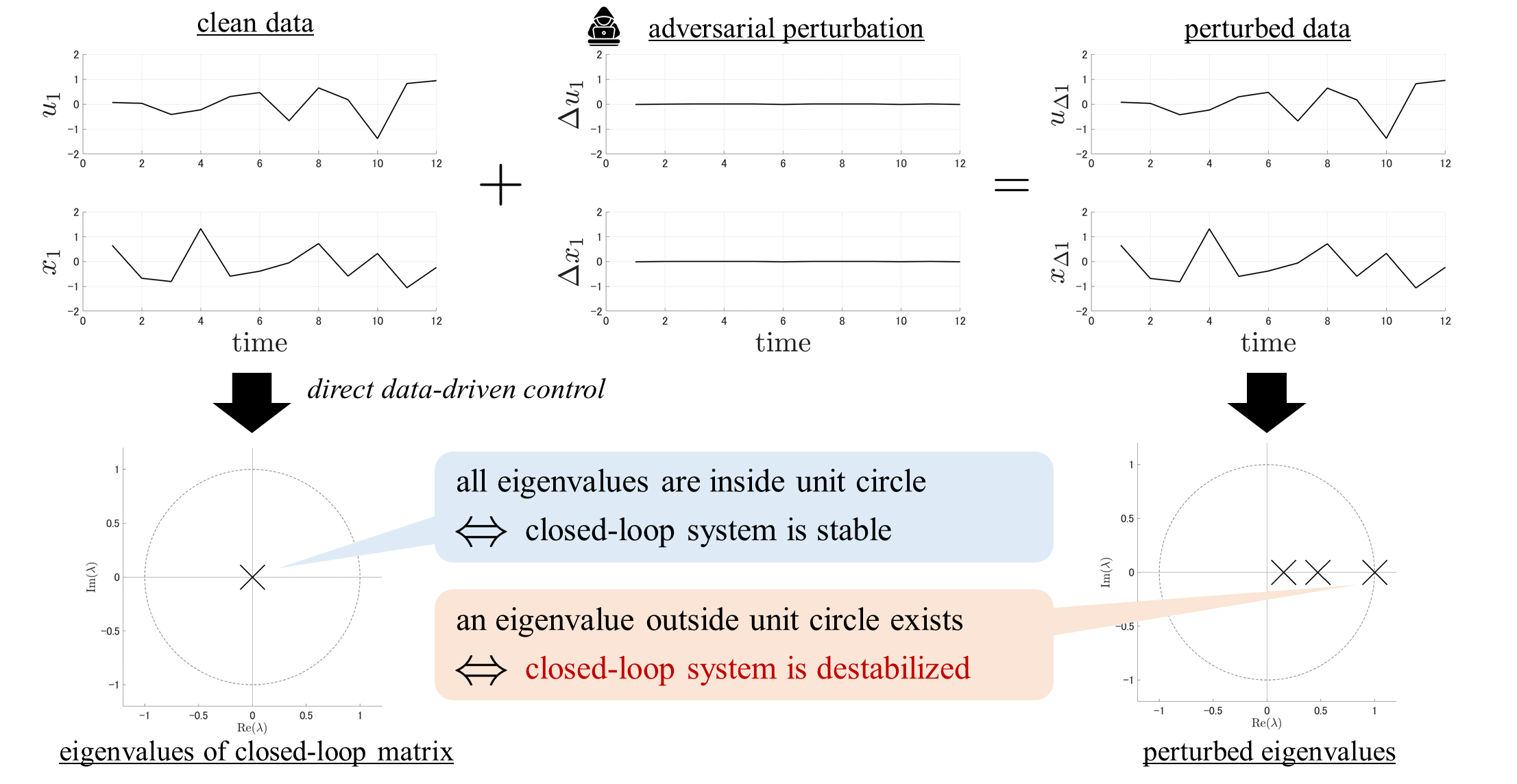}
  \caption{
  Visualization of the adversarial attack.
  In the upper half, the clean data, the adversarial perturbation, and the perturbed signal of the first input and state signals are depicted.
  In the lower half, the eigenvalues of the closed-loop matrix with the controller designed using clean data and the perturbed data.
  Note that the second and third input and state signals are also perturbed but its illustration is omitted for clarity.
  }
  \label{fig:vuln_ins}
\end{figure*}

First of all, we illustrate the adversarial attack.
Fig.~\ref{fig:vuln_ins} presents a visualization of the proposed adversarial attack applied to TT using I-DGSM with $\epsilon=0.005$ where the regularization parameter is set to $\gamma=0.0001$.
In the upper half, the clean data, the adversarial perturbation, and the perturbed signal of the first input and state signals are depicted.
In the lower left, the eigenvalues of the resulting closed-loop system with the clean data are depicted, while in the lower right, those with the perturbed data are shown.
The illustration highlights that while the system can be stabilized using clean data, resulting in eigenvalues significantly distant from the unit circle, a small yet sophisticated perturbation can render the system unstable.
The poles of the open-loop system, those of the closed-loop system with the clean data, and those of the closed-loop system with the perturbed data are summarized in Table~\ref{table:eigenvalues}.
As observed, the controller designed using clean data yields poles near the origin, indicating strong stability, whereas the adversarially perturbed data leads to a pole outside the unit circle, causing instability.
Notably, the perturbed signals appear visually similar to the clean ones, suggesting that detecting such malicious modifications through inspection alone would be challenging.
This result underscores the vulnerability of direct data-driven control methods to adversarial data manipulation.

\begin{table}[t]
\caption{Poles of open-loop system, closed-loop system with clean data, and closed-loop system with perturbed data.}
\label{table:eigenvalues}
\centering
\begin{tabular}{lccc}
\toprule
 & \multicolumn{3}{c}{{\bf Poles}}\\
 \cmidrule(lr){2-4}
 Open-loop & 0.9700 & 0.9600 & 0.9000\\
 Closed-loop (clean data) & 0.0012 & 0.0003 & 0.0001\\
 Closed-loop (perturbed data) & 1.0016 & 0.4723 & 0.1550\\
\bottomrule
\end{tabular}
\end{table}

\subsection{Attack Impact Evaluation}

\begin{figure*}[t]
  \centering
  \includegraphics[width=0.98\linewidth]{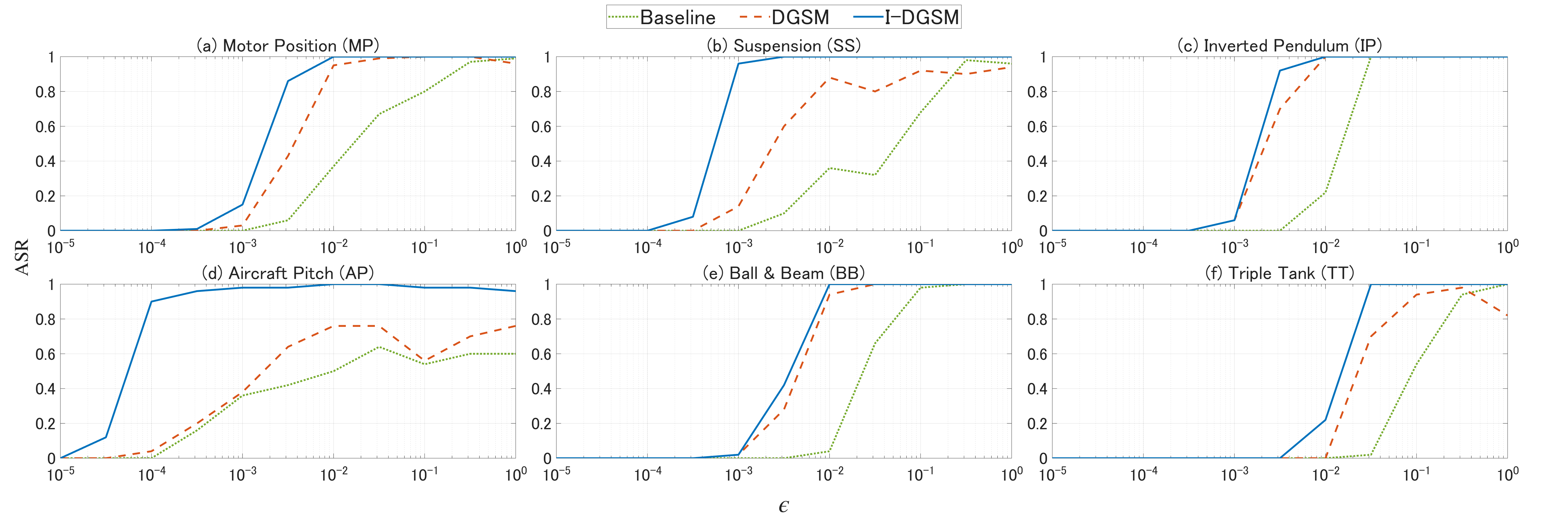}
  \caption{
  ASR over 50 trials as a function of the perturbation size~$\epsilon$ (shown on the horizontal axis in log scale) for the three perturbation generation methods (baseline, DGSM, and I-DGSM) across six benchmark systems.
  Subfigure titles indicate the specific system.
  }
  \label{fig:attack_impact}
\end{figure*}

Next, we evaluate the attack impact of the three perturbation generation algorithms.
Fig.~\ref{fig:attack_impact} shows ASR over 50 trials as a function of the perturbation size $\epsilon$ for six benchmark systems, using the baseline, DGSM, and I-DGSM algorithms, where the regularization parameter is set to $\gamma=1$.
The dotted, dashed, and solid lines represent the baseline, DGSM, and I-DGSM, respectively.
The horizontal axis is shown on a logarithmic scale to highlight differences across a wide range of perturbation magnitudes.

This result reveals several consistent patterns.
First, across all benchmark systems, I-DGSM consistently induces instability with the smallest perturbation sizes, demonstrating its effectiveness in identifying the most vulnerable directions.
DGSM requires slightly larger perturbations to cause instability but still outperforms the baseline method in most cases.
The baseline approach, in contrast, requires significantly larger perturbations to achieve similar effects.
These trends are in line with expectations, as I-DGSM incorporates optimization that explicitly targets the most destabilizing perturbations.

Note that the performance gap between I-DGSM and the baseline is substantial.
In MP, IP, BB, and TT, I-DGSM achieves a comparable ASR using perturbations that are approximately $10^{-1}$ times smaller.
This gap increases to around $10^{-2}$ in SS and AP.
These results underscore the efficiency and precision of I-DGSM in identifying minimal but impactful perturbations for destabilizing the system.

Furthermore, to verify the effectiveness of the rank complement proposed in Sec.~\ref{subsec:rankcomp}, Table~\ref{tab:defense_reg} reports the base-10 logarithm of the condition numbers of the coefficient matrix in~\eqref{eq:lin_SDP_aug} under I-DGSM with $\epsilon=10^{-3}$ (mean $\pm$ standard deviation over all iterations).
As shown in Table~\ref{tab:defense_reg}, the condition numbers for MP, SS, IP, BB, and TT remain at moderate orders of magnitude, indicating acceptable numerical conditioning.
In contrast, AP exhibits a substantially larger condition number.
While the effectiveness of the proposed method is demonstrated separately in Fig.~\ref{fig:attack_impact}, the poor conditioning under AP is consistent with the unstable behavior observed within the range $[10^{-3},10^0]$ suggesting that additional stabilization mechanisms, such as regularization, may be beneficial for more advanced attacks.

\setlength{\tabcolsep}{3pt}
\begin{table}[t]
\caption{Log condition numbers of the coefficient matrix of~\eqref{eq:lin_SDP_aug}.}
\label{tab:defense_reg}
\centering
\begin{tabular}{@{}r r r r r r@{}}
\toprule
 MP & SS & IP & AP & BB & TT\\
 \midrule
 $3.8\pm0.1$
& $4.6\pm0.5$
& $4.9\pm0.0$
& $8.5\pm0.0$
& $4.7\pm0.0$
& $1.8\pm0.0$\\
\bottomrule
\end{tabular} 
\end{table}

\setlength{\tabcolsep}{6pt}

\subsection{Computational Cost Evaluation}

We here evaluate the effectiveness of the proposed gradient computation method.
Table~\ref{tab:comp} compares the computation times in seconds (mean $\pm$ standard deviation over 10 trials) required to evaluate the gradient $\nabla_D\rho(A+B\mathcal{K}(D))$ under different differentiation methods:
NumDiff refers to numerical differentiation where the entire gradient is computed using the central difference method~\cite[Chap.~4]{Burden2015Numerical}.
ImpDiffN refers to implicit differentiation with numerical derivatives where the overall gradient is computed via the implicit differentiation as described in Algorithm~2 but the required derivatives in~\eqref{eq:deriv_spec_D} and~\eqref{eq:lin_SDP_aug} are evaluated numerically using central difference method.
ImpDiffA refers to implicit differentiation with analytical derivatives where both the gradient and the necessary intermediate derivatives are computed using the closed-form expressions provided in Section~\ref{subsec:further_efficient}.

\begin{table}[t]
\caption{Computation time (in seconds) for evaluating\\ gradient $\nabla_D\rho(A + B\mathcal{K}(D))$.}
\label{tab:comp}
\centering
\begin{tabular}{@{}l r r r@{}}
\toprule
 & NumDiff & ImpDiffN & ImpDiffA \\
\midrule
MP & 41.874$\pm$2.234 & 0.613$\pm$0.036 & 0.380$\pm$0.025\\
SS & 99.730$\pm$6.668 & 1.373$\pm$0.051 & 0.470$\pm$0.044\\
IP & 83.401$\pm$1.531 & 1.125$\pm$0.062 & 0.470$\pm$0.026\\
AP & 54.695$\pm$1.868 & 0.761$\pm$0.096 & 0.479$\pm$0.035\\
BB & 81.427$\pm$1.435 & 1.135$\pm$0.061 & 0.473$\pm$0.028\\ 
TT & 82.322$\pm$5.130 & 1.087$\pm$0.106 & 0.411$\pm$0.038\\
\bottomrule
\end{tabular}
\end{table}

A substantial improvement is observed when moving from NumDiff to ImpDiffN.
Across all six instances, the computation time is reduced by factors ranging from approximately 68$\times$ (MP: 41.87~s to 0.61~s) to 75$\times$ (TT: 82.32~s to 1.09~s), with other cases such as IP, SP, AP, and BB also showing speedups of 74$\times$, 73$\times$, 72$\times$, and 72$\times$, respectively.
This dramatic reduction confirms that even when numerical differentiation is still used to compute intermediate quantities, implicit differentiation offers a significant computational advantage by avoiding redundant gradient evaluations.
An additional improvement is achieved by switching from numerical to analytical evaluation of the derivatives, i.e., from ImpDiffN to ImpDiffA.
The resulting speedups range from about 1.6$\times$ (MP: 0.61~s to 0.38~s) to 2.9$\times$ (SS: 1.37~s to 0.47~s), with similar gains for other systems (IP: 2.4$\times$, AP: 1.6$\times$, BB: 2.4$\times$, TT: 2.6$\times$).
This further highlights the importance of analytical formulations in maximizing the efficiency of gradient computations, particularly in settings where repeated evaluations for large-scale systems are required.

\subsection{Effectiveness of Defense Strategies}
\label{subsec:effect_defense}

Subsequently, we evaluate the effectiveness of the two proposed defense strategies.

\subsubsection{Regularization}

\begin{figure}[t]
  \centering
  \includegraphics[width=0.98\linewidth]{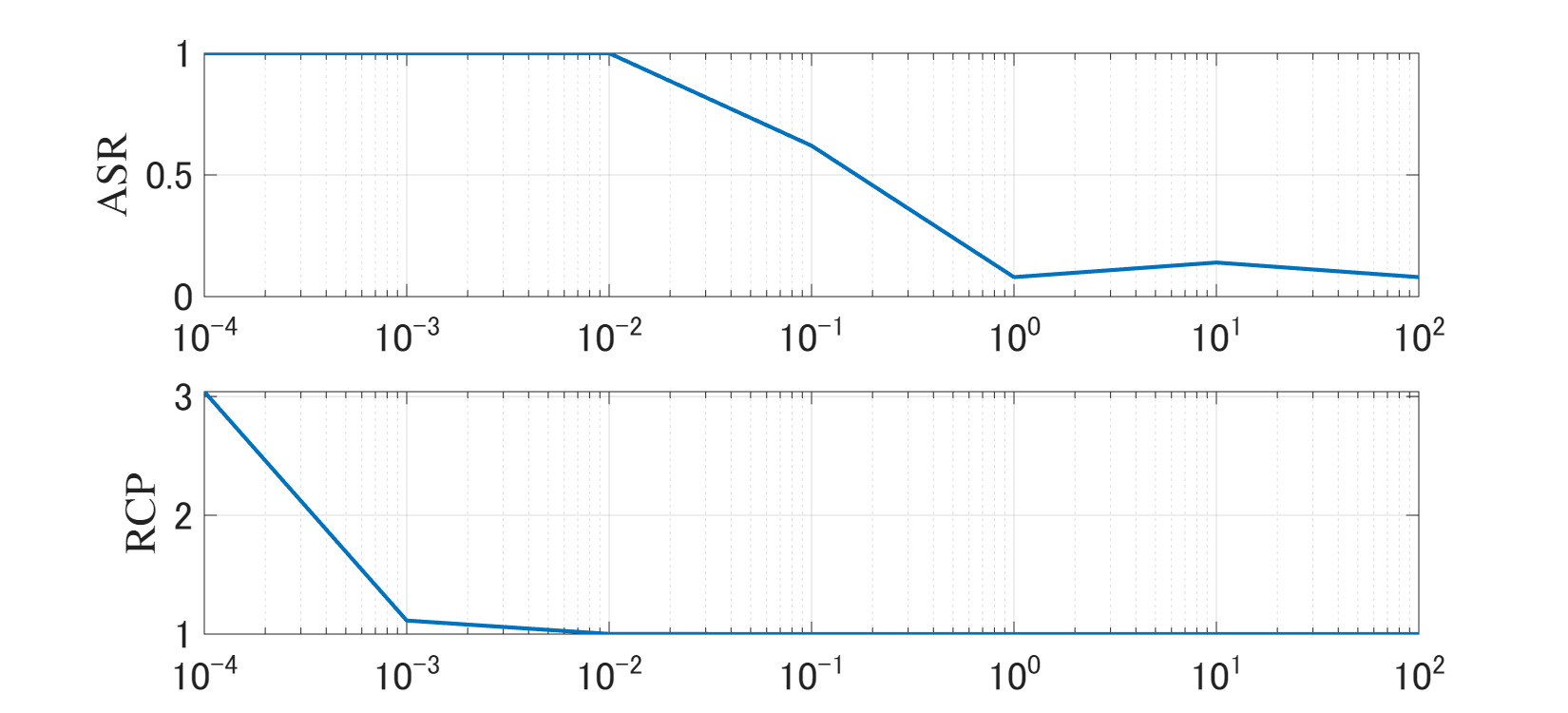}
  \caption{
  ASR and mean of RCP $\mc{J}(K_{\rm reg})/\mc{J}(K_{\rm LQR})$ over 50 trials as a function of the regularization parameter~$\gamma$ (shown on the horizontal axis in log scale) for IP with the perturbation size $\epsilon=0.001$.
  }
  \label{fig:defense_reg}
\end{figure}

Since the appropriate range of regularization parameters can vary depending on the system, we begin with a detailed analysis of the IP system as a representative example.
The top panel of Fig.~\ref{fig:defense_reg} shows ASR over 50 trials as a function of the regularization parameter $\gamma$ for the IP system.
We can observe a clear trend that the ASR decreases as $\gamma$ increases.
Further, to assess the impact of regularization on control performance, we also evaluate the averaged relative control performance (RCP) $\mc{J}(K_{\rm reg})/\mc{J}(K_{\rm LQR})$, where $K_{\rm reg}$ and $K_{\rm LQR}$ denote the controller designed by~\eqref{eq:sdp_form} and the ideal LQR optimal controller, respectively.
Since the SDP~\eqref{eq:sdp_form} yields the exact solution under clean data, we add random noise (uniformly distributed in $[-\epsilon, \epsilon]$), with which the designed controller maintains the stability, to simulate perturbed data.
The bottom panel of Fig.~\ref{fig:defense_reg} illustrates that the control performance improves as the regularization becomes stronger under noisy data.
Specifically, RCP decreases from 6.22 at $\gamma = 10^{-4}$ to 1.00 at $\gamma = 10^2$.
Table~\ref{tab:defense_reg} summarizes these trends for the other systems, where we define $\epsilon_{\rm e} \coloneqq \log_{10}\epsilon$ and $\gamma_{\rm e} \coloneqq \log_{10}\gamma$.
In all cases, the ASR decreases with increasing $\gamma$, while the RCP remains largely stable or improves, confirming the robustness benefits of regularization.
These results collectively demonstrate the effectiveness of the proposed regularization-based defense method.

\begin{table}[t]
\caption{Effectiveness of regularization-based defense strategy.}
\label{tab:defense_reg}
\centering
\begin{tabular}{@{}l r r r r@{}}
\toprule
 & $\epsilon_{\rm e}$ & Varied $\gamma_{\rm e}$ & Varied ASR & Varied RCP\\
 \midrule
MP & $-3$ & $-2\to +0$ & $1.00 \to 0.14$ & $1.00 \to 1.00$\\
SS & $-4$ & $-4\to -2$& $0.98 \to 0.00$ & $1.00 \to 1.00$\\
IP & $-3$ & $-2\to +0$ & $1.00 \to 0.04$ & $1.00 \to 1.00$\\
AP & $-5$ & $-5\to-3$ & $0.94 \to 0.00$ & $1.00 \to 1.00$\\
BB & $-3$ & $-3\to-1$ & $1.00 \to 0.00$ & $1.03\to 1.00$\\
TT & $-2$ & $-5\to-3$ & $0.86 \to 0.22$ & $1.05\to 1.00$\\
\bottomrule
\end{tabular}  
\end{table}

\subsubsection{Robust Data-driven Control}

We next evaluate the effectiveness of the robust data-driven control strategy.
In addition to ASR for stability metric, we compute the RCP, defined as $\mc{J}(K_{\rm rob}) / \mc{J}(K_{\rm LQR})$, where $K_{\rm rob}$ is the controller obtained from the robust data-driven formulation~\eqref{eq:SDP_robust} while minimizing $J_{\rm LQR}$.
We test the ASR and RCP for various perturbation magnitudes $\epsilon_{\rm e}$.
Table~\ref{tab:defense_rob} reports the results, where entries marked as N/A indicate that the SDP~\eqref{eq:SDP_robust} is infeasible.
We first note that the stability is maintained (i.e., ASR is zero) for every instance as long as the feasibility is secured, although ASR values are omitted from the table for notational simplicity.
In terms of RCP, the results show that the relative performance remains close to $1$ when $\epsilon$ is on the order of $10^{-6}$, indicating minimal performance degradation.
However, as $\epsilon$ increases, the RCP grows significantly, and the problem becomes infeasible for $\epsilon$ around $10^{-3}$ to $10^{-2}$, which is a range where the attack achieves high ASR in most systems as shown in Fig.~\ref{fig:attack_impact}.
In comparison with Table~\ref{tab:defense_reg}, these results suggest that it is less effective in preserving control performance than the regularization-based defense strategy while the robust data-driven control strategy provides formal stability guarantees.

\begin{table}[t]
\caption{RCP $\mc{J}(K_{\rm rob})/\mc{J}(K_{\rm LQR})$ with robust data-driven control for various $\epsilon_{\rm e}$.}
\label{tab:defense_rob}
\centering
\begin{tabular}{@{}l r r r r r@{}}
\toprule
  & \multicolumn{5}{c}{$\mc{J}(K_{\rm rob})/\mc{J}(K_{\rm LQR})$ for} \vspace{1mm}\\
  &$\epsilon_{\rm e}=-6$ & $-5$ & $-4$ & $-3$ & $-2$\\
   \cmidrule{1-6}
MP & 1.05 & 1.42 & N/A & N/A & N/A\\
SS & 1.01 & 1.24 & 5.62 & N/A & N/A\\
IP & 1.00 & 1.01 & 1.38 & N/A & N/A\\
AP & 1.20 & 3.21 & N/A & N/A & N/A\\
BB & 1.00 & 1.01 & 1.17 & 2.86 & N/A\\
TT & 1.00 & 1.00 & 1.01 & N/A & N/A\\
\bottomrule
\end{tabular}
\end{table}

\subsection{Transferability Evaluation}

Finally, we investigate transferability across data, where the adversary does not have access to the dataset $D$ used for controller design.
Instead, the adversary generates a hypothetical dataset $D_{\rm hyp}$ based on synthetic input-state trajectories.
Specifically, the hypothetical input is sampled as $u_{{\rm hyp},t} \sim \mc{N}(0, I_m)$, with an initial state $x_{{\rm hyp},0} \sim \mc{N}(0, I_n)$, and the resulting state trajectory $x_{{\rm hyp},t}$ is generated for $t = 1, \ldots, T$ by simulating the system dynamics.

\begin{figure*}[t]
  \centering
  \includegraphics[width=0.98\linewidth]{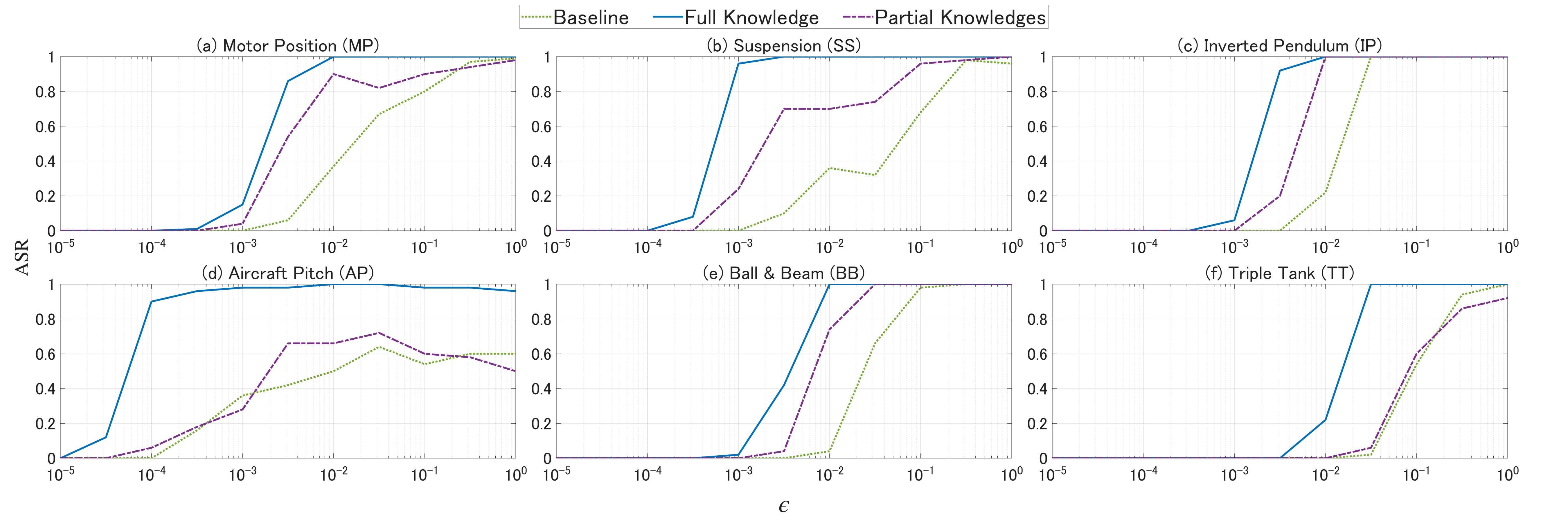}
  \caption{
  ASR over 50 trials as a function of the perturbation size~$\epsilon$ (shown on the horizontal axis in log scale) for the three cases (baseline, full knowledge, and partial knowledge) across six benchmark systems.
  Subfigure titles indicate the specific system.
  }
  \label{fig:transferability}
\end{figure*}

Fig.~\ref{fig:transferability} illustrates ASR over 50 trials as a function of the perturbation magnitude $\epsilon$ for three scenarios (baseline, full knowledge, and partial knowledge) across six benchmark systems with $\gamma=1$.
From these graphs, we observe that the partial knowledge attacks typically require perturbations with magnitudes about $10^1$ to $10^2$ times larger than those in the full knowledge case to achieve comparable ASR.
Nevertheless, the partial knowledge attacks are more effective than the baseline in most cases.
These results indicate that data transferability is moderate under this threat model and suggest that maintaining the confidentiality of the training data can serve as a practical defense against adversarial perturbations in control systems.

\section{Conclusion}
\label{sec:conc}

In this study, we have revealed the vulnerability of direct data-driven control to adversarial destabilization attacks.
We have proposed effective perturbation generation algorithms, namely DGSM and I-DGSM, tailored for control systems.
To enable practical attack evaluation, we have developed an efficient gradient computation method based on implicit differentiation through the KKT conditions of the underlying semidefinite program.
For defense, we have introduced two proactive strategies: a regularization-based approach and a robust data-driven control method.
Extensive numerical experiments have demonstrated the severity of the attacks, the computational efficiency of our methods, and the effectiveness of the proposed defenses.

Future work includes theoretical analysis of data transferability remains important directions.
Additionally, resolving the issue on differentiability of the SDP solution is also included.
Furthermore, developing more sophisticated attack strategies that explicitly account for online anomaly detection mechanisms, including residual-based detectors, within data-driven control frameworks is crucial.
Finally, this study assumes that the adversary has access to the clean data and can therefore identify the system dynamics via system identification.
If such access is imperfect due to noises, a reasonable adversary may instead rely on an approximate model obtained from partial data or prior knowledge, in which case the efficiency of the attack depends on the accuracy of the model.
Developing practical attack strategies under imperfect model knowledge is an important direction.

\begin{ack}
This work was supported in part by JSPS KAKENHI Grant Number JP24K17296 and JST-ASPIRE Program Grant Number JPMJAP2402.
\end{ack}

\appendix
\section{Appendix}
\subsection{Basic Rules of Matrix Calculus}
\label{app:mat_der}

Following the basic matrix calculus~\cite{Magnus2019Matrix}, we define the derivative of a matrix-valued function $F(X)=[f_1\ \cdots\ f_m]\in\mathbb{R}^{n\times m}$ with respect to the matrix variable $X=[x_1\ \cdots\ x_q]\in\mathbb{R}^{p\times q}$ by
\begin{equation}\label{eq:matrix_derivative}
 \dfrac{dF}{dX} := \dfrac{d {\rm vec}(F)}{d {\rm vec}(X)} = \left[
 \begin{array}{ccc}
    \dfrac{d f_1}{d x_1} & \cdots & \dfrac{d f_1}{d x_q}  \\
     \vdots & & \vdots\\
     \dfrac{d f_m}{d x_1} & \cdots & \dfrac{d f_m}{d x_q}
 \end{array}
 \right] \in \mathbb{R}^{nm \times pq},
\end{equation}
where $df_i/dx_j\in\mathbb{R}^{n\times p}$ denotes the corresponding Jacobian matrix.
We define the partial derivative $\partial F/\partial X$ in a similar manner.
We define the gradient of a scalar function $f(X)$ with respect to the matrix variable $X\in\mathbb{R}^{p\times q}$ by $\nabla_X f={\rm vec}^{-1}_{p\times q}((df/dX)^{\sf T})\in\mathbb{R}^{p\times q},$ where ${\rm vec}^{-1}_{p\times q}:\mathbb{R}^{pq}\to\mathbb{R}^{p\times q}$ denotes the inverse of the vectorization operator.
Based on the definition~\eqref{eq:matrix_derivative}, we have the chain rule for $H(X)=G(F(X))$ as
\begin{equation}
 dH/dX = \left. dG/dY\right|_{Y=F(X)} dF/dX
\end{equation}
and the product rule for $H(X)=F(X)G(X)$ with $F(X)\in\mathbb{R}^{n\times m},G(X)\in\mathbb{R}^{m\times \ell}$ as
\begin{equation}
 dH/dX=(G^{\sf T}\otimes I_n)dF/dX + (I_\ell\otimes F)dG/dX.
\end{equation}
Table~\ref{tab:formulae} exhibits useful formulae given $F(X)\in\mathbb{R}^{n\times m}$ with $X\in\mathbb{R}^{p\times q}$~\cite[Chap.~9]{Magnus2019Matrix}.

\begin{table}[t]
\caption{Matrix derivative formulae.}
\label{tab:formulae}
\centering
\begin{tabular}{@{}l l@{}}
\toprule
$F$ & $dF/dX$ \\
\midrule
$X^{\sf T}$ & $C_{p,q}$\\
$X^{-1}$ & $-X^{-{\sf T}}\otimes X^{-1}$\\
$AXB$ & $B^{\sf T}\otimes A$\\
${\rm vec}(X)^{\sf T}$ & $I_{pq}$\\
$\trace(AXB)$ & $({\rm vec}(BA)^{\sf T})^{\sf T}$\\
$\trace(X^{\sf T}AX)$ & $({\rm vec}(A+A^{\sf T})X)^{\sf T}$\\
$A\otimes X$ with $A\in\mathbb{R}^{r\times s}$ & $(I_s\otimes C_{q,r} \otimes I_p)({\rm vec}(A)\otimes I_{pq})$\\
$X\otimes A$ with $A\in\mathbb{R}^{r\times s}$ & $(I_q\otimes C_{s,p} \otimes I_r)(I_{pq}\otimes {\rm vec}(A))$\\
\bottomrule
\end{tabular}
\end{table}

\subsection{Memory-efficient Derivative Computation}
\label{app:mem_eff}

The adjoint-based method~\cite{Bell2008Algorithmic} supported by AD~\cite{Paszke2017Automatic} aims at computing quantities of the form $w^{\sf T}dL/dD$ for a given row vector $w^{\sf T}$ where $L$ satisfies the constraint $G(L,D)=0$.
Note that the sensitivity follows implicit-differentiation relation
\[
 \dfrac{\partial G}{\partial L}\dfrac{dL}{dD} + \dfrac{\partial G}{\partial D} = 0.
\]
Directly constructing the full Jacobian $dG/dL$ is computationally infeasible if its size is large.
The adjoint method avoids this by introducing an adjoint variable $\xi$ satisfying the adjoint equation
\[
 -\xi^{\sf T}\dfrac{\partial G}{\partial L} = w^{\sf T}.
\]
Once $\xi$ is obtained, one can compute the desired product without forming $dL/dD$:
\[
 w^{\sf T}\dfrac{dL}{dD} = \xi^{\sf T}\dfrac{\partial G}{\partial D}.
\]
Crucially, solving the adjoint equation only requires the ability to compute Jacobian-vector and vector-Jacobian products involving $G$, i.e., operations of the form $\partial G/\partial L v$ and $w^{\sf T} \partial G/\partial L$, which can be computed efficiently by AD in forward or reverse mode.
This structure makes the problem ideally suited to Krylov-subspace iterateve solvers, such as LSMR or GMRES (Generalized Minimal Residual Method).

\subsection{Proofs}
\label{app:proofs}

\begin{proof}
\emph{Proof of Proposition~\ref{prop:rank-deficient}:}
Let us focus on the fourth block row of $H$, namely
\begin{equation}\label{eq:Hfourth}
\begin{aligned}
 & [0_{(3n+m)^2,nT}\ 0_{(3n+m)^2,m^2}\ 0_{(3n+m)^2,1}\ I_{(3n+m)^2}]H\\
 &=
\begin{bmatrix}
    \partial G_4/\partial L & \partial G_4/\partial S & \partial G_4/ \partial \Lambda
\end{bmatrix}.
\end{aligned}
\end{equation}
Recall that $G_4=\Lambda-\Lambda^{\sf T}$.
Since the first diagonal component of $G_4$ is always zero, the first row of~\eqref{eq:Hfourth} is also zero given any variables.
Similarly, since the other diagonal components are always zero, and the corresponding rows are also zero.
Because the dimension of $G_4$ is $(3n+m)\times(3n+m)$, $H$ contains $3n+m$ zero row vectors at least.
Thus, the rank of $H$ is less than
$nT+m^2+1+(3n+m)^2-(3n+m)<nT+m^2+(3n+m)^2$, which the number of columns.
This leads to the claim.
\hfill $\square$
\end{proof}

\begin{proof}\emph{Proof of Proposition~\ref{prop:trace_reduction}:}
The sufficiency is obvious.
We here show the necessity.
Consider the Cholesky decomposition~\cite[Fact~10.10.42]{Bernstein2018Scalar} $S_i=L_iL_i^{\sf T}$ with lower triangular matrices $L_i$ for $i=1,2$.
Because $S_1\succeq 0$, we have $L_2^{\sf T}S_1L_2\succeq0$.
Thus all eigenvalues of $L_2^{\sf T}S_1L_2$ are nonnegative.
Now $\trace(L_2^{\sf T}S_1L_2)=\trace(S_1L_2L_2^{\sf T})=\trace(S_1S_2)=0$.
Because trace is equal to the sum of all eigenvalues, all eigenvalues of $L_2^{\sf T}S_1L_2$ are zero, which implies $L_2^{\sf T}S_1L_2$ is nilpotent.
Since $L_2^{\sf T}S_1L_2$ is symmetric, we have $L_2^{\sf T}S_1L_2=0$.
Then $L_2^{\sf T}S_1L_2=L_2^{\sf T}L_1L_1^{\sf T}L_2=(L_1^{\sf T}L_2)^{\sf T}L_1^{\sf T}L_2=0$, which means $L_1^{\sf T}L_2=0$.
Therefore, $S_1S_2=L_1L_1^{\sf T}L_2L_2^{\sf T}=0$.
\hfill $\square$
\end{proof}

\begin{proof}\emph{Proof of Proposition~\ref{prop:der_FK}:}
In terms of $d\hat{\rho}/dK$, since $\|A_{\rm cl}\|_{\rm F}^2={\rm tr}(A_{\rm cl}A_{\rm cl}^{\sf T})$, the cain rule and Table~\ref{tab:formulae} lead to the analytic form.

Next, from the product rule, the chain rule, and Table~\ref{tab:formulae}, we have
$\partial F_K/\partial L=((XL)^{-{\sf T}} \otimes I_m)(I_n \otimes U)- ((XL)^{-{\sf T}} \otimes F_K)(I_n \otimes X)$ and $\partial F_K/\partial D = ((XL)^{-{\sf T}} \otimes I_m)(L^{\sf T} \otimes E_U)-((XL)^{-{\sf T}} \otimes F_K)(L^{\sf T} \otimes E_X),$
which lead to the analytic forms.
\hfill $\square$
\end{proof}

\begin{proof}\emph{Proof of Proposition~\ref{prop:der_Lag}:}
Since $\|\Pi L\|_{\rm F}^2=\trace(L^{\sf T} \Pi^{\sf T}\Pi L)=\trace(L^{\sf T} \Pi L),$
from Table~\ref{tab:formulae} we have
\begin{equation}
\begin{aligned}
& \partial J/\partial L = \vecT{X^{\sf T}Q^{\sf T}}+2\gamma \vecT{\Pi L},\\
& \partial J/\partial S = \vecT{I_m},
\end{aligned}
\end{equation}
which lead to~\eqref{eq:LLLS}.
Furthermore, because $F=\bar{E}_1 F_1\bar{E}_1^{\sf T} + \bar{E}_2 F_2 \bar{E}_2^{\sf T}$, we have~\eqref{eq:FF1F2}.
Finally, considering the decomposition
\begin{equation}\label{eq:decomp_F1}
\begin{aligned}
 F_1 &= \begin{bmatrix}
     I_m\\0_{n,m}
 \end{bmatrix}
 S
 \begin{bmatrix}
     I_m\ 0_{m,n}
 \end{bmatrix}+
 \begin{bmatrix}
     I_m\\0_{n,m}
 \end{bmatrix}
 VUL
 \begin{bmatrix}
     0_{n,m}\ I_n
 \end{bmatrix}\\
&\quad + \begin{bmatrix}
    0_{m,n}\\I_n
\end{bmatrix}
(VUL)^{\sf T}
\begin{bmatrix}
    I_m\ 0_{m,n}
\end{bmatrix}+
\begin{bmatrix}
    0_{m,n}\\ I_n
\end{bmatrix}XL
\begin{bmatrix}
    0_{n,m}\ I_n
\end{bmatrix}
\end{aligned}
\end{equation}
and that for $F_2$ in a similar manner, we obtain~\eqref{eq:der_Lag}.
\hfill $\square$
\end{proof}

\begin{proof}\emph{Proof of Proposition~\ref{prop:ana_form_coe_mat}:}
Using the decompositions of $F_1$, $F_2$, $\partial F_1/\partial L$, and $\partial F_2/\partial L$ as in~\eqref{eq:decomp_F1}, together with the formulae in Table~\ref{tab:formulae}, we obtain the desired derivatives.
For the derivative $d\Pi/dD$, we apply the identity $\iGam^{\dagger} = \iGam^{\sf T}(\iGam\iGam^{\sf T})^{-1}$ since $\iGam$ is a full-row rank matrix.
\hfill $\square$
\end{proof}

\begin{proof}\emph{Proof of Proposition~\ref{prop:defense_regularization}:}
Since $Z=[B\ A]\iGam$ and $\iGam$ is full-row rank, we have $[B\ A]=Z\iGam^\dagger$ and $Z=[B\ A](\iGam_\Delta-\Delta_{\iGam})$.
Thus, from~\eqref{eq:KI},
$Z\GDel=A+B\KDel-Z\iGam^\dagger\Delta_\iGam\GDel,$
which leads to the closed-loop matrix expression $A+B\KDel=Z(I+\iGam^\dagger \Delta_\iGam)\GDel$.
Hence, the stability condition can be characterized by the Lyapunov inequality ${\rm Lyap}(P)\prec 0$ for some $P\succ 0$ where
\begin{equation}
    {\rm Lyap}(P)\coloneqq Z(I+\iGam^\dagger \Delta_\iGam)\GDel P \GDel^{\sf T}(I+\iGam^\dagger \Delta_\iGam)^{\sf T}Z^{\sf T}-P.
\end{equation}

Let us take $\PDel$ as a candidate that satisfies the Lyapunov inequality.
Note that, from the constraint in~\eqref{eq:prob_reg1}, $\PDel$ satisfies $Z_\Delta\MDel Z_\Delta^{\sf T}-\PDel+I_n\preceq 0$.
Now we define $\Psi\coloneqq {\rm Lyap}(\PDel)-Z_\Delta\MDel Z_\Delta^{\sf T}+\PDel$.
Using basic algebra, we have
\begin{equation}
\begin{aligned}
 \Psi = &  -Z\MDel \Delta_Z^{\sf T} - \Delta_Z\MDel Z^{\sf T} - \Delta_Z\MDel\Delta_Z^{\sf T}+Z\iGam^\dagger\Delta_\iGam \MDel Z^{\sf T}\\
  & + Z\MDel(\iGam^\dagger \Delta_\iGam)^{\sf T} Z^{\sf T} + Z\iGam^\dagger \Delta_\iGam \MDel (\iGam^\dagger \Delta_\iGam)^{\sf T} Z^{\sf T}.\\ 
\end{aligned}
\end{equation}
Because every operator norm is submultiplicative and $\|\iGam^\dagger\|_2=\sigma_{\rm min}(\iGam)^{-1}$, we have
\begin{equation}
\begin{aligned}
 \|\Psi\|_2 &  \leq \|Z\|_2^2\|\MDel\|_2 (\|\Delta_Z\|_2^2 / \|Z\|_2^2 + 2\|\Delta_Z\|_2/\|Z\|_2\\
  & \quad \quad+ \|\Delta_\iGam\|_2^2/\sigma_{\rm min}(\iGam)^2 + 2 \|\Delta_\iGam\|_2/\sigma_{\rm min}(\iGam))\\
  & \leq 2\|Z\|_2^2\|\MDel\|_2(\mu^{-2}+2\mu^{-1}).
\end{aligned}
\end{equation}
Therefore, if $\mu$ satisfies the inequality, we have $\|\Psi\|_2< 1$, which is equivalent to $\Psi\prec I$.
This implies that ${\rm Lyap}(\PDel)\prec Z_\Delta\MDel Z_\Delta^{\sf T}-\PDel+I\prec 0,$ which leads to the claim.
\hfill $\square$
\end{proof}

\begin{proof}\emph{Proof of Proposition~\ref{prop:defense_robust}:}
We first consider the perturbation-free case, i.e., $\Delta = 0$.
In this case, it follows that $D_\Delta = D$ and $Z_\Delta = AX_\Delta + BU_\Delta$.
Define $K = F_K(L, D_\Delta)$, $P = X_\Delta L$, and $G = LP^{-1}$, where $P$ is nonsingular due to the constraint $F_2 \succeq 0$.
By applying the Schur complement to the condition $F(L, S, D_\Delta) \succeq 0$, it can be verified that the triplet $(K, P, G)$ satisfies the constraints in~\eqref{eq:prob_ori}.
Therefore, the nominal performance is guaranteed by $\mathcal{J}(K) = \operatorname{trace}(QX_\Delta L) + \operatorname{trace}(S) \leq J_{\rm LQR}$.

Next, consider the case where $\|\Delta\|_{\max} \leq \epsilon$.
We have $\|\Delta\|_2 \leq \|\Delta\|_{\mathrm{F}} \leq \sqrt{(2n + m)T} \|\Delta\|_{\max} \leq \sqrt{(2n + m)T} \epsilon,$
which implies $\Delta \Delta^{\mathsf{T}} \preceq \epsilon^2 (2n + m)T I_{2n + m}$.
This is equivalent to the quadratic matrix inequality $[I\ \Delta]{\rm diag}(\epsilon^2(2n+m)TI_{2n+m},-I_T)[I\ \Delta]^{\sf T}\preceq 0$.
According to Theorem~3.8 in~\cite{Kaminaga2025Data}, the third, fourth, and fifth constraints in~\eqref{eq:SDP_robust} constitute a sufficient condition for the stability of the closed-loop system.
\end{proof}

\bibliographystyle{plain}        
\bibliography{sshrrefs}

@article{Martinez2021Software,
  title={Software supply chain attacks, a threat to global cybersecurity: {SolarWinds'} case study},
  author={Mart{\'\i}nez, Jeferson and Dur{\'a}n, Javier M},
  journal={International Journal of Safety and Security Engineering},
  volume={11},
  number={5},
  pages={537--545},
  year={2021}
}

@article{Bell2008Algorithmic,
  title={Algorithmic differentiation of implicit functions and optimal values},
  author={Bell, Bradley M and Burke, James V},
  journal={Proc. Advances in Automatic Differentiation},
  pages={67--77},
  year={2008},
}

@article{Paszke2017Automatic,
  title={Automatic Differentiation in {PyTorch}},
  author={Paszke, Adam and Gross, Sam and Chintala, Soumith and Chanan, Gregory and Yang, Edward and DeVito, Zachary and Lin, Zeming and Desmaison, Alban and Antiga, Luca and Lerer, Adam},
  journal={Proc. Advances in Neural Information Processing Systems (NeurIPS)},
  year={2017}
}

@book{Bonnans2000Perturbation,
  title={Perturbation Analysis of Optimization Problems},
  author={Bonnans, J. Fr{\'e}d{\'e}ric and Shapiro, Alexander},
  year={2000},
  publisher={Springer Science \& Business Media}
}

@article{Bellon2025Parametric,
  title={Parametric semidefinite programming: {G}eometry of the trajectory of solutions},
  author={Bellon, Antonio and Henrion, Didier and Kungurtsev, Vyacheslav and Mare{\v{c}}ek, Jakub},
  journal={Mathematics of Operations Research},
  volume={50},
  number={1},
  pages={410--430},
  year={2025},
}

@article{Kozyakin2009Accuracy,
  title={On accuracy of approximation of the spectral radius by the {Gelfand} formula},
  author={Kozyakin, Victor},
  journal={Linear Algebra and its Applications},
  volume={431},
  number={11},
  pages={2134--2141},
  year={2009},
}

@article{Zhang2009Smoothing,
  title={Smoothing projected gradient method and its application to stochastic linear complementarity problems},
  author={Zhang, Chao and Chen, Xiaojun},
  journal={SIAM Journal on Optimization},
  volume={20},
  number={2},
  pages={627--649},
  year={2009},
}

@article{Anand2025Analysis,
  title={Analysis and Mitigation of Data injection Attacks against Data-Driven Control},
  author={Anand, Sribalaji C},
  journal={arXiv},
  note = {[Online]. Available: \url{https://arxiv.org/pdf/2504.17347}},
  year={2025}
}

@misc{Sasahara2025Adversraial_GitHub,
  author = {Hampei Sasahara},
  note = {\url{https://github.com/HampeiSasahara/adv-des-attack-dddc/releases/tag/v2.0}},
  doi = {10.5281/zenodo.16204191},
  year = {2025},
}

@article{Schwarting2018Planning,
  title={Planning and decision-making for autonomous vehicles},
  author={Schwarting, Wilko and Alonso-Mora, Javier and Rus, Daniela},
  journal={Annual Review of Control, Robotics, and Autonomous Systems},
  volume={1},
  number={1},
  pages={187--210},
  year={2018},
}

@article{O2022Data,
  title={Data-driven predictive control with improved performance using segmented trajectories},
  author={O’Dwyer, Edward and Kerrigan, Eric C and Falugi, Paola and Zagorowska, Marta and Shah, Nilay},
  journal={IEEE Trans. Control Syst. Technol.},
  volume={31},
  number={3},
  pages={1355--1365},
  year={2022},
}

@article{Schmitt2023Data,
  title={Data-driven predictive control with online adaption: {Application} to a fuel cell system},
  author={Schmitt, Lukas and Beerwerth, Julius and Bahr, Matthias and Abel, Dirk},
  journal={IEEE Trans. Control Syst. Technol.},
  volume={32},
  number={1},
  pages={61--72},
  year={2023},
}

@article{Wang2021Stop,
  title={Stop-and-go: {Exploring} backdoor attacks on deep reinforcement learning-based traffic congestion control systems},
  author={Wang, Yue and Sarkar, Esha and Li, Wenqing and Maniatakos, Michail and Jabari, Saif Eddin},
  journal={IEEE Trans. Inf. Forensics Security},
  volume={16},
  pages={4772--4787},
  year={2021},
  publisher={IEEE}
}

@article{Deka2022Dynamically,
  title={Dynamically computing adversarial perturbations for recurrent neural networks},
  author={Deka, Shankar A and Stipanovi{\'c}, Du{\v{s}}an M and Tomlin, Claire J},
  journal={IEEE Trans. Control Syst. Technol.},
  volume={30},
  number={6},
  pages={2615--2629},
  year={2022},
}

@inproceedings{Song2023Discovering,
  title={Discovering adversarial driving maneuvers against autonomous vehicles},
  author={Song, Ruoyu and Ozmen, Muslum Ozgur and Kim, Hyungsub and Muller, Raymond and Celik, Z Berkay and Bianchi, Antonio},
  booktitle={Proc. 32nd USENIX Security Symposium (USENIX Security 23)},
  pages={2957--2974},
  year={2023}
}

@inproceedings{Eykholt2018Robust,
  title={Robust physical-world attacks on deep learning visual classification},
  author={Eykholt, Kevin and Evtimov, Ivan and Fernandes, Earlence and Li, Bo and Rahmati, Amir and Xiao, Chaowei and Prakash, Atul and Kohno, Tadayoshi and Song, Dawn},
  booktitle={Proc. IEEE Conference on Computer Vision and Pattern Recognition},
  pages={1625--1634},
  year={2018}
}

@inproceedings{Garcia2017Hey,
  title={Hey, My Malware Knows Physics! {Attacking PLCs} with Physical Model Aware Rootkit.},
  author={Garcia, Luis and Brasser, Ferdinand and Cintuglu, Mehmet Hazar and Sadeghi, Ahmad-Reza and Mohammed, Osama A and Zonouz, Saman A},
  booktitle={Proc. Network and Distributed System Security (NDSS) Symposium},
  year={2017}
}

@inproceedings{Biggio2012Poisoning,
author = {Biggio, Battista and Nelson, Blaine and Laskov, Pavel},
title = {Poisoning attacks against support vector machines},
year = {2012},
booktitle = {Proc. 29th International Coference on Machine Learning},
pages = {1467–1474},
}

@inproceedings{Yu2023Poisoning,
  title={Poisoning Attacks Against Data-Driven Predictive Control},
  author={Yu, Yue and Zhao, Ruihan and Chinchali, Sandeep and Topcu, Ufuk},
  booktitle={Proc. American Control Conference (ACC)},
  pages={545--550},
  year={2023},
}

@article{Ikezaki2024Poisoning,
  title={Poisoning attack on {VIMT} and its adverse effect},
  author={Ikezaki, Taichi and Kaneko, Osamu and Sawada, Kenji and Fujita, Junya},
  journal={Artificial Life and Robotics},
  volume={29},
  number={1},
  pages={168--176},
  year={2024},
}

@inproceedings{Kaminaga2024Adversarial,
  title={Adversarial Attack using Projected Gradient Method to Direct Data-Driven Control},
  author={Kaminaga, Taira and Sasahara, Hampei},
  booktitle={Proc. IEEE Conference on Control Technology and Applications (CCTA)},
  pages={236--241},
  year={2024},
}

@article{Campi2002Virtual,
  title={Virtual reference feedback tuning: {A} direct method for the design of feedback controllers},
  author={Campi, Marco C and Lecchini, Andrea and Savaresi, Sergio M},
  journal={Automatica},
  volume={38},
  number={8},
  pages={1337--1346},
  year={2002},
}

@article{Kaneko2013Data,
  title={Data-driven controller tuning: {FRIT} approach},
  author={Kaneko, Osamu},
  journal={IFAC Proceedings Volumes},
  volume={46},
  number={11},
  pages={326--336},
  year={2013},
}

@article{Yuan2019Adversarial,
  title={Adversarial examples: {Attacks} and defenses for deep learning},
  author={Yuan, Xiaoyong and He, Pan and Zhu, Qile and Li, Xiaolin},
  journal={IEEE Transactions on Neural Networks and Learning Systems},
  volume={30},
  number={9},
  pages={2805--2824},
  year={2019},
}

@misc{Matlab2024,
  author = {{MathWorks}},
  title = {{MATLAB version 24.2.0 (R2024b)}},
  howpublished = {Software},
  address = {Natick, MA, USA},
  year = {2024},
  note = {\url{https://www.mathworks.com}}
}

@book{Bernstein2018Scalar,
  title={Scalar, Vector, and Matrix Mathematics: {Theory,} Facts, and Formulas-revised and Expanded Edition},
  author={Bernstein, Dennis},
  year={2018},
  publisher={Princeton University Press}
}

@article{Duvenaud2020Deep,
  title={Deep implicit layers tutorial-neural {ODEs}, deep equilibirum models, and beyond},
  author={Duvenaud, David and Kolter, J Zico and Johnson, Matthew},
  journal={Neural Information Processing Systems Tutorial},
  year={2020}
}

@book{Boyd2004Convex,
  title={Convex Optimization},
  author={Boyd, Stephen and Vandenberghe, Lieven},
  year={2004},
  publisher={Cambridge University Press}
}

@article{Xu2024Revisiting,
  title={Revisiting implicit differentiation for learning problems in optimal control},
  author={Xu, Ming and Molloy, Timothy L and Gould, Stephen},
  journal={Proc. Advances in Neural Information Processing Systems (NeurIPS)},
  volume={36},
  year={2024}
}

@inproceedings{Amos2017Optnet,
  title={Optnet: {Differentiable} optimization as a layer in neural networks},
  author={Amos, Brandon and Kolter, J Zico},
  booktitle={Proc. International Conference on Machine Learning (ICML)},
  pages={136--145},
  year={2017},
}

@article{Blondel2022Efficient,
  title={Efficient and modular implicit differentiation},
  author={Blondel, Mathieu and Berthet, Quentin and Cuturi, Marco and Frostig, Roy and Hoyer, Stephan and Llinares-L{\'o}pez, Felipe and Pedregosa, Fabian and Vert, Jean-Philippe},
  journal={Proc. Advances in Neural Information Processing Systems (NeurIPS)},
  volume={35},
  pages={5230--5242},
  year={2022}
}

@book{Magnus2019Matrix,
  title={Matrix Differential Calculus with Applications in Statistics and Econometrics},
  author={J. R. Magnus and H. Neudecker},
  year={2019},
  publisher={John Wiley \& Sons},
}

@misc{Dawn2025Tilbury,
  title={Control Tutorials for {MATLAB and Simulink}},
  author={Dawn Tilbury and Bill Messner},
  year={2026},
  note={Accessed: 8th April, 2026, [Online.] Available: \url{https://ctms.engin.umich.edu/CTMS/index.php?aux=Home}},
}

@book{William1999Control,
  title={Control Tutorials for {MATLAB} and Simulink: {A} Web-Based Approach},
  author={Messner, William and Tilbury, Dawn},
  year={1999},
  publisher={Prentice Hall}
}

@article{Kaminaga2025Data,
  title={Data informativity under data perturbation},
  author={Taira Kaminaga and Hampei Sasahara},
  journal={arXiv preprint},
  year={2025},
  note = {[Online]. Available: \url{https://arxiv.org/pdf/2505.01641}},
}

@INPROCEEDINGS{Kaminaga2025DataACC,
author={Taira Kaminaga and Hampei Sasahara},
booktitle={Proc. 2025 American Control Conference},
title={Data informativity for quadratic stabilization under data perturbation},
year={2025},
volume={},
number={},
pages={},}

@article{Tian2021Joint,
  title={Joint adversarial example and false data injection attacks for state estimation in power systems},
  author={Tian, Jiwei and Wang, Buhong and Wang, Zhen and Cao, Kunrui and Li, Jing and Ozay, Mete},
  journal={IEEE Trans. Cybern.},
  volume={52},
  number={12},
  pages={13699--13713},
  year={2021},
}

@article{Dorfler2023Data,
  title={Data-Driven Control: {P}art One of Two: {A} Special Issue Sampling from a Vast and Dynamic Landscape},
  author={D{\"o}rfler, Florian},
  journal={IEEE Control Systems Magazine},
  volume={43},
  number={5},
  pages={24--27},
  year={2023},
}

@inproceedings{Sasahara2023Adversarial,
  title={Adversarial Attacks to Direct Data-driven Control for Destabilization},
  author={Sasahara, Hampei},
  booktitle={Proc. IEEE 62nd Conference on Decision and Control (CDC)},
  year={2023},
  pages={7094--7099},
}

@book{Burden2015Numerical,
  title={Numerical Analysis},
  author={Burden, Richard L and Faires, J Douglas and Burden, Annette M},
  year={2015},
  edition={10th},
  publisher={Cengage learning}
}

@article{Van2020Data,
  title={Data informativity: {A} new perspective on data-driven analysis and control},
  author={Van Waarde, Henk J and Eising, Jaap and Trentelman, Harry L and Camlibel, M Kanat},
  journal={IEEE Trans. Autom. Control},
  volume={65},
  number={11},
  pages={4753--4768},
  year={2020},
}

@book{Chen2012Optimal,
  title={Optimal Sampled-data Control Systems},
  author={Chen, Tongwen and Francis, Bruce A},
  year={2012},
  publisher={Springer}
}

@article{Dorfler2023Bridging,
  title={Bridging direct \& indirect data-driven control formulations via regularizations and relaxations},
  author={D{\"o}rfler, Florian and Coulson, Jeremy and Markovsky, Ivan},
  journal={IEEE Trans. Autom. Control},
  volume={68},
  number={2},
  pages={883-897},
  year={2023},
}

@inproceedings{Carlini2016Hidden,
  title={Hidden voice commands},
  author={Carlini, Nicholas and Mishra, Pratyush and Vaidya, Tavish and Zhang, Yuankai and Sherr, Micah and Shields, Clay and Wagner, David and Zhou, Wenchao},
  booktitle={25th USENIX security symposium (USENIX security 16)},
  pages={513--530},
  year={2016}
}

@inproceedings{Bruna2014Intriguing,
title	= {Intriguing properties of neural networks },
author	= {Joan Bruna and Christian Szegedy and Ilya Sutskever and Ian Goodfellow and Wojciech Zaremba and Rob Fergus and Dumitru Erhan},
year	= {2014},
booktitle	= {Proc. International Conference on Learning Representations (ICLR)}
}

@inproceedings{Goodfellow2015Explaining,
title	= {Explaining and Harnessing Adversarial Examples},
author	= {Ian Goodfellow and Jonathon Shlens and Christian Szegedy},
year	= {2015},
booktitle	= {Proc. International Conference on Learning Representations (ICLR)}
}

@article{De2019Formulas,
  title={Formulas for data-driven control: Stabilization, optimality, and robustness},
  author={De Persis, Claudio and Tesi, Pietro},
  journal={IEEE Trans. Autom. Control},
  volume={65},
  number={3},
  pages={909--924},
  year={2019},
}

@article{Willems2005Note,
  title={A note on persistency of excitation},
  author={Willems, Jan C and Rapisarda, Paolo and Markovsky, Ivan and De Moor, Bart LM},
  journal={Systems \& Control Letters},
  volume={54},
  number={4},
  pages={325--329},
  year={2005},
}

@book{Lewis2013Reinforcement,
  title={Reinforcement Learning and Approximate Dynamic Programming for Feedback Control},
  author={Lewis, Frank L and Liu, Derong},
  year={2013},
  publisher={John Wiley \& Sons}
}

@inproceedings{Dorfler2022On,
  title={On the Role of regularization in direct data-driven {LQR} control},
  author={D\"{o}rfler, Florian and Tesi, Pietro and De Persis, Claudio},
  booktitle={Proc. IEEE 61st Conference on Decision and Control (CDC)},
  pages={1091--1098},
  year={2022},
}

@book{Blanke2016Diagnosis,
author = {Mogens Blanke and Michel Kinnaert and Jan Lunze and Marcel Staroswiecki},
title = {Diagnosis and Fault-Tolerant Control},
publisher = {Springer},
year = {2016},
edition = {3},
}

@article{Milosevic2019Estimating,
  title={Estimating the impact of cyber-attack strategies for stochastic networked control systems},
  author={Milo{\v{s}}evi{\'c}, Jezdimir and Sandberg, Henrik and Johansson, Karl Henrik},
  journal={IEEE Trans. Control Netw. Syst.},
  volume={7},
  number={2},
  pages={747--757},
  year={2019},
}
\end{document}